\documentclass[]{aa}
\usepackage{graphicx}
\usepackage{txfonts}

\usepackage{natbib}
\bibpunct{(}{)}{;}{a}{}{,} 
\usepackage{color}
\usepackage[colorlinks=true,linkcolor=black,citecolor=blue,urlcolor=blue,draft=false]{hyperref}
\makeatletter
\renewcommand*\aa@pageof{, page \thepage{} of \pageref*{LastPage}}
\makeatother
\begin{document} 

   \title{Interpreting high spatial resolution line observations of planet-forming disks with gaps and rings: The case of HD 163296}


  \author{Ch. Rab\inst{1,5}  
  \and I. Kamp\inst{1}
  \and C. Dominik\inst{2}
  \and C. Ginski\inst{2,3}   
  \and G. A. Muro-Arena\inst{2}
  \and W.-F. Thi\inst{5}
  \and L. B. F. M. Waters\inst{4,2}
  \and \mbox{P. Woitke\inst{6,7}}}

  \institute{Kapteyn Astronomical Institute, University of Groningen, P.O. Box 800, 9700 AV Groningen, The Netherlands \email{rab@astro.rug.nl}
  \and Anton Pannekoek Institute for Astronomy, University of Amsterdam, Science Park 904, 1098 XH Amsterdam, The Netherlands
  \and Leiden Observatory, Leiden University, PO Box 9513, 2300 RA Leiden, The Netherlands
  \and  SRON Netherlands Institute for Space Research, Sorbonnelaan 2, 3584 CA Utrecht, The Netherlands
  \and Max-Planck-Institut f\"ur extraterrestrische Physik, Giessenbachstrasse 1, 85748 Garching, Germany
  \and SUPA, School of Physics \& Astronomy, University of St. Andrews, North Haugh, St. Andrews KY16 9SS, UK
  \and Centre for Exoplanet Science, University of St. Andrews, North Haugh, St. Andrews, KY16 9SS, UK
  } 

   \date{Received June 22 2020 / Accepted August 13 2020}

 
  \abstract
   {Spatially resolved continuum observations of planet-forming disks show prominent ring and gap structures in their dust distribution. However, the picture from gas observations is much less clear and constraints on the radial gas density structure (i.e. gas gaps) remain rare and uncertain. 
   }
   {We want to investigate the importance of thermo-chemical processes for the interpretation of high-spatial-resolution gas observations of planet-forming disks and their impact on the derived gas properties.}
   {We applied the radiation thermo-chemical disk code P{\tiny RO}D{\tiny I}M{\tiny O} (PROtoplanetary DIsk MOdel) to model the dust and gas disk of \mbox{HD 163296 self-consistently}, using the DSHARP (Disk Substructure at High Angular Resolution) gas and dust observations. With this model we investigated the impact of dust gaps and gas gaps on the observables and the derived gas properties, considering chemistry, and heating and cooling processes.}
   {We find distinct peaks in the radial line intensity profiles of the CO line data of \mbox{\object{HD 163296}} at the location of the dust gaps. Our model indicates that those peaks are not only a consequence of a gas temperature increase within the gaps but are mainly caused by the absorption of line emission from the back side of the disk by the dust rings. For two of the three prominent dust gaps in \mbox{HD 163296}, we find that thermo-chemical effects are negligible for deriving density gradients via measurements of the rotation velocity. However, for the gap with the highest dust depletion, the temperature gradient can be dominant and needs to be considered to derive accurate gas density profiles.}
   {Self-consistent gas and dust thermo-chemical modelling in combination with high-quality observations of multiple molecules are necessary to accurately derive gas gap depths and shapes. This is crucial to determine the origin of gaps and rings in planet-forming disks and to improve the mass estimates of forming planets if they are the cause of the gap.}
\titlerunning{Gaps and rings in HD 163296}
\maketitle
%
\section{Introduction}
With modern telescopes such as the Atacama Large Millimeter Array (ALMA) at millimetre-wavelengths or the SPHERE (Spectro-Polarimetric High-contrast Exoplanet REsearch) instrument at the Very Large Telescope (VLT/SPHERE) at optical wavelengths, it has become possible to produce spatially resolved images of planet-forming disks surrounding young low and intermediate mass stars down to spatial scales of a few astronomical units. These observations have revealed structures such as spiral arms, vortices, and, most prominently, gaps and rings in dust continuum observations. Large high-spatial-resolution ALMA surveys indicate that dust gaps and rings are common in observed planet-forming disks \citep{Andrews2018,Long2018}. Compared to ALMA, rings and gaps are not as frequently observed in scattered light images, which might be caused by observational biases. VLT/SPHERE scattered light observations show gaps and rings mostly in bright and extended disks \citep{Garufi2018}, furthermore shallow gaps are likely  to be harder to detect in the optical compared to the millimetre regime due to optical depth effects. However, considering those factors, \citet{Garufi2018} concluded that scattered light observations also indicate that rings and gaps in disks are a normality.

The most common explanation of the observed dust gaps are forming planets that carve gaps and cause a pile up of larger dust particles in pressure bumps, which are observed as mostly azimuthally symmetric ring structures \citep[e.g.][]{Pinilla2012,Dipierro2015,Zhang2018}. However, other interpretations are also possible such as enhanced pebble growth at molecular ice lines \citep[e.g.][]{Pinilla2017,Zhang2015b,Owen2020}, dust evolution and inefficient fragmentation of dust particles (gaps in scattered light, \citealt{Birnstiel2015}), or magnetized disks \citep[e.g.][]{Flock2015,Bethune2017,Riols2020}. Larger sample studies of continuum observations do not exclude any of those scenarios, however, the ice-line scenario is unlikely to be a universal mechanism as most observed gaps and rings are not associated with the expected locations of molecular ice lines \citep[e.g.][]{Huang2018,Long2018,vanderMarel2019}. The various gap formation theories differ especially in their predictions for the gas. For example in the planet scenario a dust gap is associated with a gas gap, but this is not the case for the ice line scenario or for dust gaps produced at the edge of dead zones in magnetized disks. A single planet might also produce multiple gaps in the gas and dust as shown for example by \citet{Zhang2018}. However, the models of \citet{Ziampras2020} indicate that radiative effects can suppress multiple gap formation (i.e. each gap requires a planet). Accurate gas and dust surface density measurements of planet-forming disks that allow us to estimate the gap depths and shapes are required to provide constraints for the various gap formation scenarios. 

In a different manner to the dust, gas observations do not provide a clear picture of the presence of gaps and rings in the gas disk. One reason for this is that spectral line observations are more expensive in terms of observing time compared to continuum observations. Furthermore the interpretation of molecular line data is more complex due to thermo-chemical and kinematic effects. For example ring structures in molecular line observations can be purely caused by chemical effects such as the freeze-out of molecules, photo-desorption, or changing dust properties (e.g. \citealt{Oeberg2015a} molecule: DCO$^+$; \citealt{Cleeves2016b} CO; \citealt{Bergin2016} C$_2$H; \citealt{Salinas2017} DCO$^+$, N$_2$D$^+$; \citealt{Cazzoletti2018} CN;
\citealt{Qi2019} N$_2$H$^+$). This complexity allows for different interpretations of the same gas observations as for example discussed by \citet{vanderMarel2018} for the case of \mbox{HD 163296}. So far indications of the presence of gas gaps that might have been produced by planets exist for only a few disks: e.g. HL~Tau, \citealt{Yen2016}; \mbox{HD 163296}, \citealt{Isella2016,Teague2018}; AS~209, \citealt{Favre2019}; \mbox{TW Hya}, \citealt{Teague2017}.

Not included in the above list are transitional disks with their large ($r\gtrsim20\,\mathrm{au}$), strongly dust depleted (often no dust is detected at millimetre wavelengths) inner cavities. Observations of transition disks clearly show strong gas depletion (by factors of \mbox{$\gtrsim\!100$}) within their dust cavities. However, extensive studies and thermo-chemical models have shown that the gas cavity is usually smaller than the dust cavity and also that the gas is not as strongly depleted as the dust \citep[e.g.][]{Bruderer2014,Carmona2014n,vanderMarel2016,DrabekMaunder2016,Dong2017b,Boehler2018,UbeiraGabellini2019}. Although the physical conditions in the cavities are different compared to gaps in full disks (e.g. cavities are also optically thin in the dust at infrared wavelengths), the thermo-chemical models used for transitional disk studies have already addressed many aspects of dust and gas gap modelling (e.g. determining gas surface density profiles) and we use here a very similar modelling approach.

In this work we focus on the disk around the Herbig Ae/Be star HD~163296, which is of particular interest for studying gas observations as three generations of ALMA CO molecular line and continuum observations exist (e.g. \citealt{deGregorio-Monsalvo2013a,Flaherty2015,Isella2016}) that were all combined to produce the DSHARP (Disk Substructure At High Angular Resolution Project) dataset of \mbox{HD 163296} for the dust at $1.25\,\mathrm{mm}$ and the \mbox{$^{12}\mathrm{CO}\,J\!=\!2\!-\!1$} line \citep{Isella2018}. Furthermore, \mbox{HD 163296} is likely to be the only disk where indirect evidence for forming planets was reported using different observational methods. Besides the indications for planets from dust gaps in millimetre observations \citep{Isella2016,Isella2018}, scattered light images also show gaps and ring structures \citep{Grady2000,Muro-Arena2018}. The strongest constraints for ongoing planet formation in \mbox{HD 163296} come from the detection of kinematic signatures in CO spectral line observations, which are most likely caused by planets. Such signatures were found for the two dust gaps at $r\approx86$ and $r\sim145\,\mathrm{au}$ \citep{Teague2018,Teague2019,Pinte2020} but also outside the observed millimetre continuum disk at $r\sim250\,\mathrm{au}$ \citep{Pinte2018}. However, it is not yet clear if all of the dust gaps in \mbox{HD 163296} host a planet. The masses of the potential planets derived via different methods can also vary by an order of magnitude \citep[e.g.][]{Pinte2020}. 

In this work we use \mbox{HD 163296} as an example to study the impact of dust gaps and possible gas gaps on high spatial resolution spectral line observations. We focus on the DSHARP dataset as it provides the highest spatial resolution for the dust and gas emission. We model this data with a self-consistent gas and dust model for \mbox{HD 163296} using the radiation thermo-chemical disk code P{\tiny RO}D{\tiny I}M{\tiny O} (PROtoplanetary DIsk MOdel). The main purpose of this paper is not to present a detailed best fit model for \mbox{HD 163296}, but to study the impact of thermo-chemical processes on observables such as radial intensity profiles and channel maps, with the aim of quantifying the importance of thermo-chemical processes for accurate measurements of the gas surface density, kinematic signatures, and pressure gradients.  

We first describe our modelling approach and our fiducial models for the disk of \mbox{HD 163296} (Sect.~\ref{sec:methods}). Using these models we investigate the origin of radial features in the observed CO line emission (Sect.~\ref{sec:radprofs}) and discuss the impact of thermo-chemical processes on measurements of the rotation velocity and pressure gradients (Sect.~\ref{sec:vrot}). In Sect.~\ref{sec:gasgaps} we present synthetic channel maps, discuss the presence of gas gaps, compare our results to previous modelling, and discuss possible model improvements. Our conclusions are presented in Sect.~\ref{sec:conclusions}. 
\section{Methods}
\label{sec:methods}
\subsection{Radiation thermo-chemical modelling}
\label{sec:rtcmodelling}
To model the disk of \mbox{HD 163296}, we used the radiation thermo-chemical disk code P{\tiny RO}D{\tiny I}M{\tiny O}\footnote{\url{https://www.astro.rug.nl/~prodimo}} (PROtoplanetary DIsk MOdel;  \citealt{Woitke2009a,Kamp2010,Woitke2016}). P{\tiny RO}D{\tiny I}M{\tiny O} consistently solves for the dust radiative transfer, gas thermal balance, and chemistry for a given static two-dimensional dust and gas density structure. Furthermore, P{\tiny RO}D{\tiny I}M{\tiny O} provides modules to produce synthetic observables such as spectral lines \citep{Woitke2011}, spectral energy distributions (SED) \citep{Thi2011}, and images. For the chemistry we use a chemical network with 235 gas and ice phase species and 2844 chemical reactions including freeze-out of atoms, molecules and PAHs (polycyclic aromatic hydrocarbons), photo-, thermal-, and cosmic-ray desorption of ices, \mbox{X-ray} chemistry, H$_2$ formation on grains, and excited H$_2$ chemistry. This chemical network is described in detail in \citet{Kamp2017}. The chemistry is solved consistently with the heating and cooling balance for the gas temperature. Important heating and cooling processes are photo-electric heating, PAH-heating, chemical heating by exothermic reactions, X-ray heating, viscous heating,  thermal-accommodation on dust grains, and line cooling by atomic and molecular species. Further details on the implementation of the heating and cooling balance and the various processes can be found in  \citet{Woitke2009a,Woitke2011} and \citet{Aresu2011}.
\subsection{HD 163296 model}
For this paper we focus on the DSHARP \mbox{$^{12}\mathrm{CO}\,J\!=\!2\!-\!1$} spectral line and the $1.25\,\mathrm{mm}$ continuum observations. As a starting point we used an existing model from the DIANA\footnote{\url{https://dianaproject.wp.st-andrews.ac.uk}} project (DIsc ANAlysis) that matches the SED and line fluxes from near-infrared to millimetre wavelengths within a factor of about two (see \citealt{Woitke2019,Dionatos2019} for details). The model includes a separate zone for the inner disk ($r\lesssim 2.5\,\mathrm{au}$) producing a jump in the gas and dust surface density. Otherwise this model has a smooth density distribution in the gas and dust. The \mbox{HD 163296}  DIANA model was also used as a starting point by \citet{Muro-Arena2018} to model VLT/SPHERE scattered light images and the pre-DSHARP continuum images \citep{Isella2016} that already show clear signatures of dust gaps and rings in both the thermal and scattered light dust emission.

We updated the DIANA \mbox{HD 163296} model considering the new GAIA distance ($d=101\,\mathrm{pc}$), adapted stellar parameters, and the disk mass of \mbox{$\sim0.2\,M_\mathrm{\sun}$} as derived by \citet{Booth2019} from $^{13}\mathrm{C}^{17}\mathrm{O}$ observations. This disk mass is about a factor of three lower than in the original DIANA model \citep{Woitke2019} and a factor of about three higher than in the older \mbox{\mbox{HD 163296}} P{\tiny RO}D{\tiny I}M{\tiny O} model of \citet{Tilling2012}. 

The model presented here is in reasonable agreement with the spatially resolved DSHARP data (neglecting the gaps) and still matches the SED and line fluxes to a similar quality as the original DIANA model. All relevant parameters of this initial model are listed in Table~\ref{table:diskmodel} and further details are provided in Appendix~\ref{sec:diskmodel_app}. We use this model as a starting point for our models with dust and gas gaps and as a reference for a disk with a smooth density distribution. How we include gaps and rings in the model is described in Sect.~\ref{sec:method_gaps}.
\subsection{Observational dataset and synthetic observables}
\label{sec:methods_obs}
For this work we focus on the \mbox{HD 163296} DSHARP\footnote{\url{https://almascience.eso.org/almadata/lp/DSHARP}} dataset \citep{Andrews2018} that is described in detail in \citet{Isella2018}. The beam size of the dust continuum observations at 1.25mm is $0\farcs038\times0\farcs048$ (\mbox{$\sim4.3\,\mathrm{au}$}) and the rms noise is $0.023\,\mathrm{mJy/beam}$. The \mbox{$^{12}\mathrm{CO}\,J\!=\!2\!-\!1$} line observations have a beam size of $0\farcs104\times 0\farcs095$ (\mbox{$\sim 10\,\mathrm{au}$}), a channel width of $0.32\,\mathrm{km\,s^{-1}}$ (but the spectral resolution is  $0.64\,\mathrm{km\,s^{-1}}$ due to Hanning smoothing), and a rms noise per channel of $0.84\,\mathrm{mJy/beam}$ \citep{Isella2018}.
 
From this dataset we produced azimuthally averaged radial intensity profiles for the dust and the gas emission following the approach of \citet{Huang2018}. We used elliptical apertures, with the minor axis given by the disk inclination $i=46.7^\circ$ and the orientation given by the position angle of the disk $\mathrm{PA}=133.3^\circ$. The width of the aperture is about one third of the beam width. For the error bars we used three times the root mean square of the data divided by the square root of independent beams within the elliptical aperture \citep{Isella2016}. This procedure gives a nearly identical dust radial profile as derived by \citet{Huang2018}, with the difference that we neglect the azimuthally asymmetric structure observed in one of the gaps (i.e. we did not remove it as was done in \citealt{Huang2018}). For the gas we apply the same procedure on the integrated intensity (moment 0) image of the \mbox{$^{12}\mathrm{CO}\,J\!=\!2\!-\!1$} line. The resulting radial intensity profiles will be discussed in Sect.~\ref{sec:radprofs}.

To produce synthetic observables such as continuum images and spectral line cubes, we use the line radiative transfer module of P{\tiny RO}D{\tiny I}M{\tiny O} \citep{Woitke2011}. The molecular data for the line excitation calculation for the CO molecule are from the LAMDA database (Leiden Atomic and Molecular Database) \citep{Jankowski2005,Yang2010,Schoeier2005c}. For $\mathrm{C^{18}O}$ we assume a fixed $\mathrm{^{16}O/^{18}O}$ isotopologue ratio of 498.7 \citep{Scott2006}. The modelled synthetic images are convolved with a 2D kernel using the beam properties of the observations. For the spectral lines the synthetic channel maps are additionally re-gridded to the velocity spacing of the observations. This was done with the CASA software package (Common Astronomy Software Applications, \citealt{McMullin2007}) using the image functions \texttt{convolve2d} and \texttt{regrid}, respectively. For a comparison to the observations we produce synthetic radial intensity profiles in the same manner as was done for the observations.

\begin{figure}
\resizebox{\hsize}{!}{\includegraphics{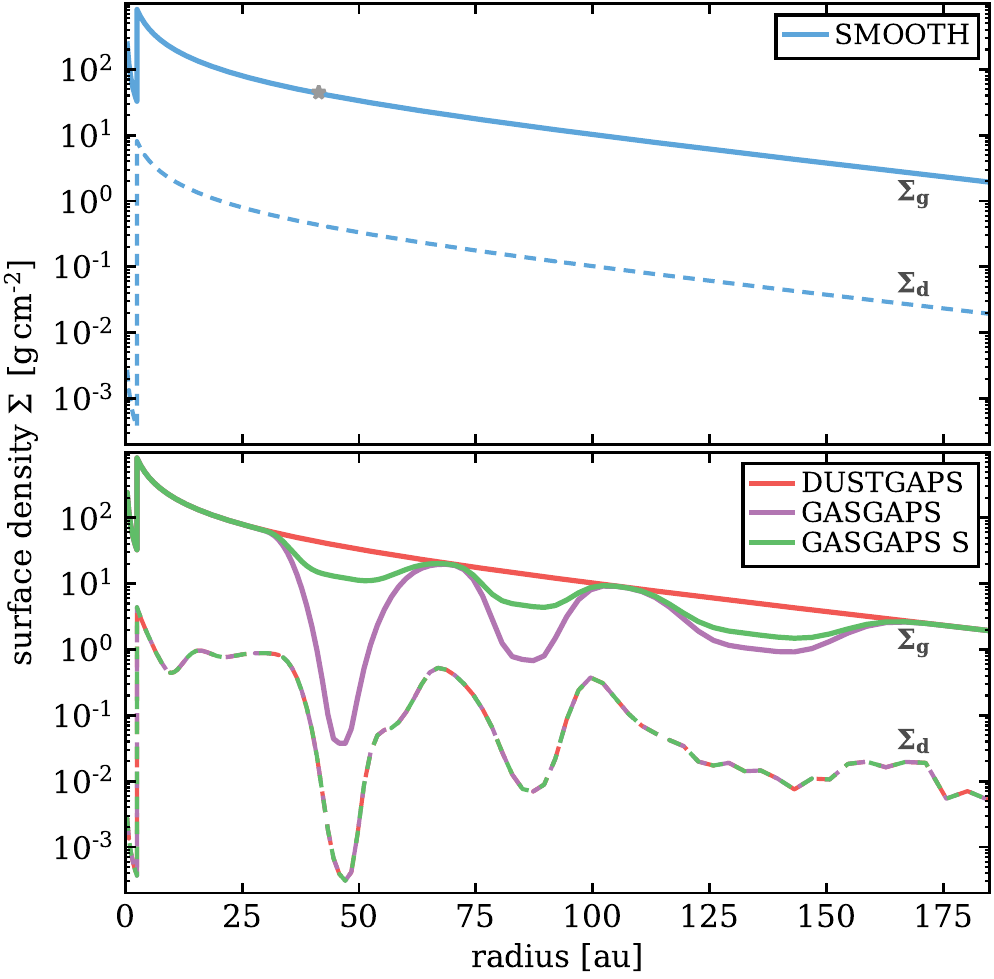}}
\caption{Gas ($\Sigma_\mathrm{g}$, solid lines) and dust ($\Sigma_\mathrm{d}$, dashed lines) surface densities for the inner 180 au of our fiducial disk models. The top panel shows the initial SMOOTH model without any gaps or rings. The grey star symbol indicates the gas surface density of $44.4\,\mathrm{g\,cm^{-2}}$ as measured by \citet{Booth2019} using $\mathrm{^{13}C^{17}O}\,J=3-2$ observations with a spatial resolution of $\approx70\,\mathrm{au}$. The bottom panel shows the models using the same fitted dust surface density profile but varying gas surface density profiles (see Sect.~\ref{sec:method_gaps}). All models are described in Table~\ref{table:models}.}
\label{fig:sdprofiles}
\end{figure}
\begin{table}
\caption{Overview of the presented models (see also Fig.~\ref{fig:sdprofiles}).}
\label{table:models}
\centering
\begin{tabular}{llp{4.5cm}}
\hline\hline
Name & Colour &Description  \\
\hline 
SMOOTH & blue &smooth gas and dust radial surface density profiles ($\Sigma_\mathrm{d}, \Sigma_\mathrm{g}$) \\
\hline 
DUSTGAPS &  red & dust gaps (fitted $\Sigma_\mathrm{d}$); smooth $\Sigma_\mathrm{g}$ \\
\noindent\parbox[t]{1.8cm}{DUSTGAPS FREEZE} & orange & same as DUSTGAPS,  but temperatures and the chemistry are fixed to the values from the SMOOTH model \\
\hline 
GASGAPS & purple & same $\Sigma_\mathrm{d}$ as DUSTGAPS but with deep gas gaps (similar depletion as for the dust) \\
GASGAPS S & green & same as GASGAPS but with shallow gas gaps \\
\hline
\end{tabular}
\end{table}
\subsection{Fiducial models: Surface densities}
\label{sec:method_gaps}
For the presentation and discussion of our results we focus on four fiducial models called SMOOTH, DUSTGAPS, \mbox{GASGAPS,} and \mbox{GASGAPS S}. The only differences between these models are the input gas ($\Sigma_\mathrm{g}$) and dust ($\Sigma_\mathrm{d}$) radial surface density profiles as shown in Fig.~\ref{fig:sdprofiles}. The model configurations are summarized in Table~\ref{table:models}. The SMOOTH model is the starting model with a smooth gas and dust radial surface density profile. 

To introduce the dust gaps in our model we follow the approach used by \citet{Pinte2016} for HL~Tau and \citet{Muro-Arena2018} for \mbox{HD 163296}. We fit the radial dust intensity profile by adapting the dust surface density via an iterative procedure but keep all other dust properties (i.e. grain size distribution) fixed. To measure the depth of the resulting dust gaps we use the approach of \citet{Huang2018}; the gap depth is then given by
\begin{equation}
    \Delta_\mathrm{gap}=\frac{\Sigma_d(r_\mathrm{ring})}{\Sigma_d(r_\mathrm{gap})}.
\end{equation}
We find gap depths of roughly \mbox{$\Delta_\mathrm{gap}\sim1700$}, 50 and 3 for the dust gaps at $r=48$ (dust gap one, DG1), 85 (DG2),  and 150 au (DG3), respectively. These numbers are roughly consistent with the results from \citet{Isella2016}. They found depletion factors of >100, 70, and 6 for their narrow dust gap model, using rectangular gaps to fit the pre-DSHARP data (six times lower spatial resolution). We note that \citet{Isella2016} measured the gap depth relative to a smooth dust density profile. The presented dust disk model provides us with an accurate model for the dust emission, which is crucial for the interpretation of the continuum subtracted line data. In this DUSTGAPS model only $\Sigma_\mathrm{d}$ is adapted whereas $\Sigma_\mathrm{g}$ is the same as in the SMOOTH model.

To produce gas gaps we apply a parameterized approach using  parameters such as gap location, gap width, and depletion factor. The shape of the gaps are a combination of Gaussian wings and a flat bottom (for details see \citealt{Oberg2020}). The gap locations and widths are chosen to be similar to the dust gaps. The depths of the gaps are given relative to $\Sigma_\mathrm{g}$ from the SMOOTH model. For the GASGAPS model we use gas depletion factors of 1000, 20, and 5 for DG1 to DG3 (i.e. similar to the dust). To study the impact of shallow gas gaps we introduce the model \mbox{GASGAPS S} where we chose a constant gas depletion factor of three for all three gas gaps. The resulting $\Sigma_\mathrm{g}$ profile for the gaps is comparable to the hydrodynamic modelling results of \citet{Teague2018}. However, we do not use their profile because we want to be consistent with our gas gap modelling approach. We note that the $\Sigma_g$ profiles around DG3 are very similar for both models with gas gaps (see Fig.~\ref{fig:sdprofiles}). We emphasize that we did not fit the gas data to produce the gas surface density profiles, they are merely used to show the impact of gas gaps on observables.

The resulting gas and dust surface density profiles are shown in Fig.~\ref{fig:sdprofiles}. For these models we self-consistently calculate the two-dimensional dust and gas temperature structure, the chemical abundances, and the synthetic observables. In Appendix~\ref{sec:diskmodel_app} we discuss further details of the models and show the resulting two-dimensional density structure, temperature structure, and far-UV radiation field for each model. The CO chemistry within the gaps is discussed in Sect.~\ref{sec:vrot} where the CO disk structure is also shown (Fig.~\ref{fig:lineorigin}).%
\begin{figure}
\resizebox{\hsize}{!}{\includegraphics{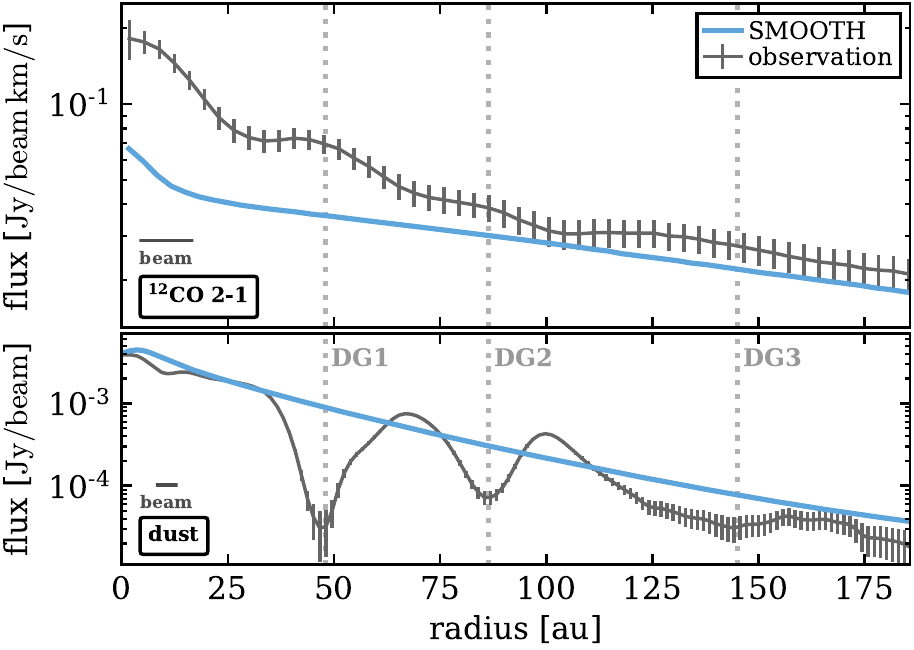}}
\caption{Azimuthally averaged radial intensity profiles for the \mbox{$^{12}\mathrm{CO}\,J\!=\!2\!-\!1$} line emission (top panel) and the  1.25 mm dust continuum emission (bottom panel). The grey solid lines with error bars show the observations. The blue solid lines show the SMOOTH model. In the bottom left of each panel we show the full-width at half-maximum (FWHM) of the beam for the gas and continuum observations, respectively. The vertical dotted grey lines indicate the location of the three prominent dust gaps.}
\label{fig:rprofs}
\end{figure}
\section{Origin of the radial features in the CO line emission}
\label{sec:radprofs}
In Fig.~\ref{fig:rprofs} we show the observed azimuthally averaged radial intensity profiles for the dust and the \mbox{$^{12}\mathrm{CO}\,J\!=\!2\!-\!1$} line in comparison to the SMOOTH model. Even though the SMOOTH model has no radial features, by construction the modelled \mbox{$^{12}\mathrm{CO}\,J\!=\!2\!-\!1$} radial profile is in good agreement with the observations. For $r>30\,\mathrm{au}$ the maximum deviation is about a factor of two but for most radii it is significantly smaller. The largest discrepancies (up to a factor of 3.2) are actually in the inner $30\,\mathrm{au}$ where the optically thick dust disk makes the interpretation of the data challenging \citep[e.g.][]{Weaver2018}. However, for our study we focus on the regions around the three prominent dust gaps indicated as DG1, DG2, and DG3 in Fig.~\ref{fig:rprofs}. 

Similarly to earlier lower spatial resolution data \citep{Isella2016,vanderMarel2018}, radial features are visible in the azimuthally averaged \mbox{$^{12}\mathrm{CO}\,J\!=\!2\!-\!1$} radial  intensity profile. However, the DSHARP data now clearly shows that the three weak peaks in the \mbox{$^{12}\mathrm{CO}\,J\!=\!2\!-\!1$} radial intensity profile coincide with the location of the three prominent dust gaps. In this section we investigate the physical origin of these radial features.%
\subsection{Impact of temperature}
\begin{figure}
\resizebox{\hsize}{!}{\includegraphics{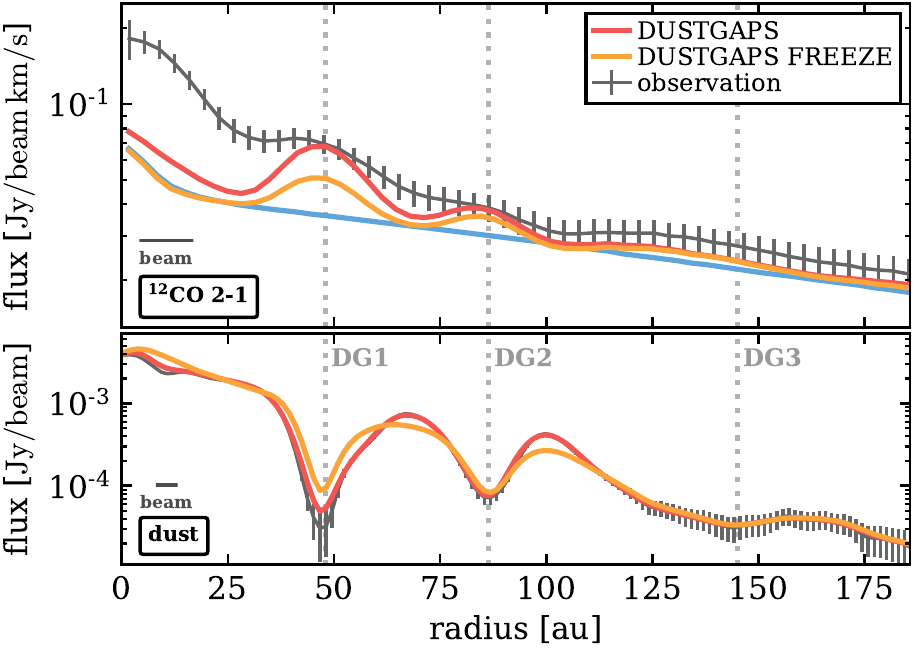}}
\caption{Same as \mbox{Fig. \ref{fig:rprofs}} but for the DUSTGAPS model and the \mbox{DUSTGAPS FREEZE} model. For the latter the gas and dust temperature and chemical structure is the same as in the SMOOTH model. In the top panel the SMOOTH model (blue solid line) is also shown for reference.}
\label{fig:rprofs_gaps}
\end{figure}

The radial gas and dust intensity profiles for the \mbox{DUSTGAPS} model are shown in Fig.~\ref{fig:rprofs_gaps}. Our dust disk model nicely fits the observed radial dust intensity profile, but also improves the match for the \mbox{$^{12}\mathrm{CO}\,J\!=\!2\!-\!1$} radial profile compared to the SMOOTH model. The modelled profile shows similar peaks at the location of the dust gaps as the observations. We emphasize that the gas density structure in the \mbox{DUSTGAPS} model is identical to the SMOOTH model. 

The most likely explanation for the peaks in the line emission of the DUSTGAPS model is actually a temperature increase within the dust gaps (see Appendix \ref{sec:diskmodel_app}). This was already postulated by the \citet{vanderMarel2018} modelling of pre-DSHARP ALMA data of \mbox{HD 163296}. They argue that even if the gas is depleted within dust gaps, the  increase in temperature will lead to an increase in the line emission at the location of dust gaps. 

To test this temperature scenario, we constructed a model called \mbox{DUSTGAPS FREEZE}, which includes the dust gaps but whose temperature and chemical structure is identical to the SMOOTH model. As seen in Fig.~\ref{fig:rprofs_gaps}, the \mbox{DUSTGAPS FREEZE} model still shows peaks in the \mbox{$^{12}\mathrm{CO}\,J\!=\!2\!-\!1$} radial profile at the location of the dust gaps, although there is no temperature increase within the dust gaps. For DG2 and DG3 the radial line intensity profiles of the DUSTGAPS and \mbox{DUSTGAPS FREEZE} model are nearly identical, whereas for DG1 the peak is significantly weaker in the \mbox{DUSTGAPS FREEZE} model. This means that for DG1 the temperature increase within the gap partly explains the increase in the line emission. However, for DG2 and DG3 this is not the case and temperature and chemical changes are not the cause of the peaks in the \mbox{$^{12}\mathrm{CO}\,J\!=\!2\!-\!1$} radial profile. 

From the bottom panel of Fig.~\ref{fig:rprofs_gaps} we see that the dust radial profile of the DUSTGAPS and \mbox{DUSTGAPS FREEZE} model are not identical. This is expected because in the \mbox{DUSTGAPS FREEZE}  model the dust temperature structure is not consistent with the dust density structure. This implies that a modelling approach with a prescribed temperature structure will lead to inaccurate results, and self-consistent dust radiative transfer modelling is required to derive accurate dust surface density profiles. However, the changes in the dust radial profile in the \mbox{DUSTGAP FREEZE} model are too weak to have a significant impact on the peaks in the \mbox{$^{12}\mathrm{CO}\,J\!=\!2\!-\!1$} radial profile. This is especially true for the outermost gap.

\citet{Isella2018} have shown that in the channel maps of the \mbox{$^{12}\mathrm{CO}\,J\!=\!2\!-\!1$} line, clear signatures of dust extinction are visible (see also Fig.~\ref{fig:masking}). As the dust rings are optically thick (or close to optically thick, \citealt{Dullemond2018}), they absorb the line emission from the back side of the disk, as this emission has to pass through the midplane of the disk and the dust rings. In the next section we explore the impact of the dust absorption process on the radial intensity profiles of spectral lines. 
\begin{figure}
\resizebox{\hsize}{!}{\includegraphics{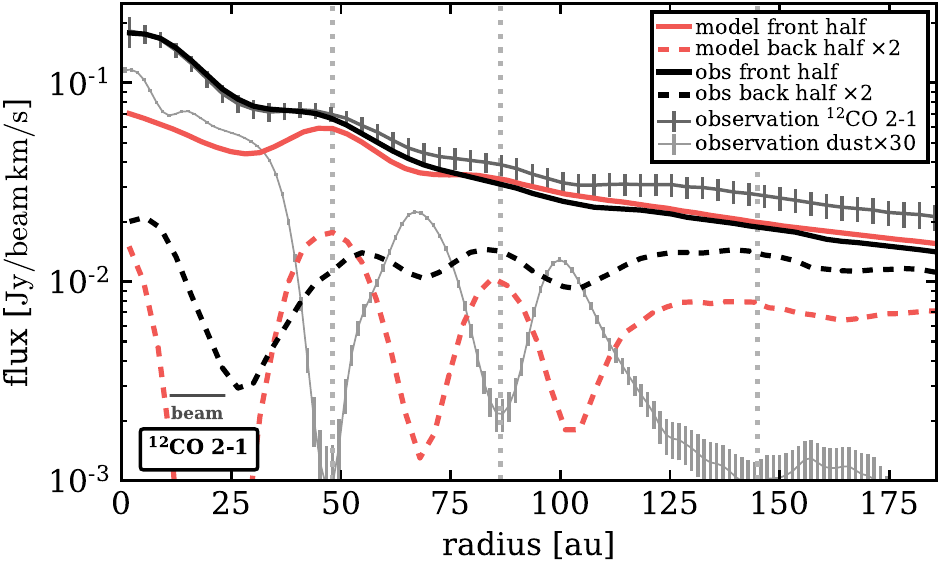}}
\caption{Radial intensity profiles for the \mbox{$^{12}\mathrm{CO}\,J\!=\!2\!-\!1$} line shown separately for the front and back side of the disk. The solid and dashed black lines show the profiles for the front and back side of the disk for the observations. The solid and dashed red lines are the front and back side profiles derived from the DUSTGAPS model. The grey solid and dashed lines with error bars show the observed full (i.e. both sides) radial profiles for the gas and dust, respectively, for reference. For easier comparison we multiplied the dust radial profile by a factor of 30 and the profiles for the back side of the disk by a factor of two.}
\label{fig:radmasks}
\end{figure}
\subsection{Contribution of the back side of the disk}
\label{sec:backside}
The observational data clearly shows the emission from the front and back side of the disk, as the disk is inclined and the \mbox{$^{12}\mathrm{CO}\,J\!=\!2\!-\!1$} emission layer is high up in the disk. Furthermore, the line emission from the back side of the disk shows radial features that are likely caused by the dust rings (see \cite{Isella2018} and Fig.~\ref{fig:masking}). We note that those features in the back side emission cannot be identified in the \mbox{pre-DSHARP} data used in \citet{Isella2016} and \citet{vanderMarel2018} due to lower spatial resolution and lower sensitivity. The high quality and spatial resolution of the DSHARP line data allows us to separate the emission from the two sides of the disk. By applying Keplerian masking, we can separate the back and front side emission and produce two sets of azimuthally averaged radial \mbox{$^{12}\mathrm{CO}\,J\!=\!2\!-\!1$} intensity profiles, one for the front and one for the back side of the disk (see Appendix \ref{sec:keplerianmasking} for details).

The observed \mbox{$^{12}\mathrm{CO}\,J\!=\!2\!-\!1$} radial profiles for the back and front side of the disk are shown in Fig.~\ref{fig:radmasks}. For DG2 and DG3 the radial features in the line emission are clearly seen in the profile for the back side of the disk, but almost vanish in the profile for the front side. For DG1 the situation is much less clear. The reason is most likely the non-perfect Keplerian masking, which becomes more difficult in the inner disk as the front and back side layers are not as well resolved due to the lower height of the emission layers.

To verify this scenario we took our DUSTGAPS model and separated the front and back half of the disk emission during the line radiative transfer step to achieve an exact separation of the two disk halves. For the resulting two line cubes we then applied the same procedure as for the full line cube to produce the radial line intensity profiles.  As seen in Fig.~\ref{fig:radmasks}, the modelled profiles show the same behaviour as the observational data. However, for the back side of the emission the radial features are much more pronounced than in the data.

Both the observational data and the model indicate that the peaks in the full \mbox{$^{12}\mathrm{CO}\,J\!=\!2\!-\!1$} radial profile are actually a consequence of the presence of dust gaps and rings. As the dust gaps are optically thin, the line emission from the back side of the disk is unaffected, whereas the rings absorb a significant fraction of the emission. This  produces the radial features in the back side emission, which are strong enough to show up in the full line radial profile. However, in the model the absorption of the line emission from the back side of the disk is too strong, as is clearly seen in Fig.~\ref{fig:radmasks}. This indicates that our dust model overestimates the extinction properties of the dust and the optical depth in the rings, despite the fact that the model nicely fits the observed thermal dust emission. 

For our model the unsettled dust opacity properties at \mbox{$\lambda=1.25\,\mathrm{mm}$} for the chosen dust size distribution are \mbox{$\kappa_\mathrm{d}^\mathrm{abs}=1.24\,\mathrm{g\,cm^{-2}}$} (absorption) and \mbox{$\kappa_\mathrm{d}^\mathrm{sca}=7.56\,\mathrm{g\,cm^{-2}}$} (scattering). Using the parameter \mbox{$\epsilon_\mathrm{d}:=\kappa_\mathrm{d}^\mathrm{abs}/(\kappa_\mathrm{d}^\mathrm{abs}+\kappa_\mathrm{d}^\mathrm{sca})$} like \citet{Isella2018}, we get $\epsilon_d=0.14$ for our dust model. \citet{Isella2018} derived values in the range of  $\epsilon_\mathrm{d}\sim0.4-0.6$ for the different dust rings, using the continuum and line data combined with dust temperature estimates from various models. This indicates that our model has a too high scattering opacity at millimetre wavelengths and therefore overestimates the total dust optical depth $\tau_\mathrm{d}^\mathrm{ext}$ within the rings. However, our dust model fits the data because scattering can reduce the observed dust emission as has been shown by \citet{Zhu2019}, \citet{Liu2019} and \citet{Ueda2020}. Consequently, the model overestimates the absorption of the back side \mbox{$^{12}\mathrm{CO}\,J\!=\!2\!-\!1$} line emission, resulting in too weak line emission at the location of the dust rings (see Fig.~\ref{fig:radmasks}). 

So far the impact of the dust disk on the back side of the gas emission has only been directly observed  for \mbox{HD 163296}  \citep{Isella2018}. However, it is likely that such an effect occurs in all disks with prominent dust gaps and rings as long as the dust rings are not optically thin. Furthermore, we can only directly see these effects in inclined disks, as for disks with low inclination the back side of the disk is hidden behind the front side of the disk. Even for the case of an optically thin line, where we observe the sum of the emission from the front and back half of the disk, the dust absorption effect will still produce features in radial intensity profiles and in channel maps. This introduces some degeneracies for the interpretation of radial gas intensity profiles as radial features can be produced by varying temperatures or the presence of optically thick dust rings, and in most cases a combination of those two. 

We show that a dust model that fits the millimetre dust image is not necessarily good enough to interpret line observations. To fix our model, it is most likely that we require a more sophisticated dust model that allows for radial variations of the dust properties (i.e. dust size distribution). This scenario is supported by the fact that our model underestimates the line emission from the back side of the disk at the location of the dust rings by factors of five to ten, but is in much better agreement (within a factor of two) at the location of the dust gaps (see Fig.~\ref{fig:radmasks}). 

As already mentioned by \citet{Isella2018}, gas lines can provide crucial constraints for the dust properties and a model as presented here is ideal to study this in more detail. However, this is beyond the scope of this paper, but will be explored in future work where we also include an improved dust model.
\begin{figure*}
\centering
\includegraphics{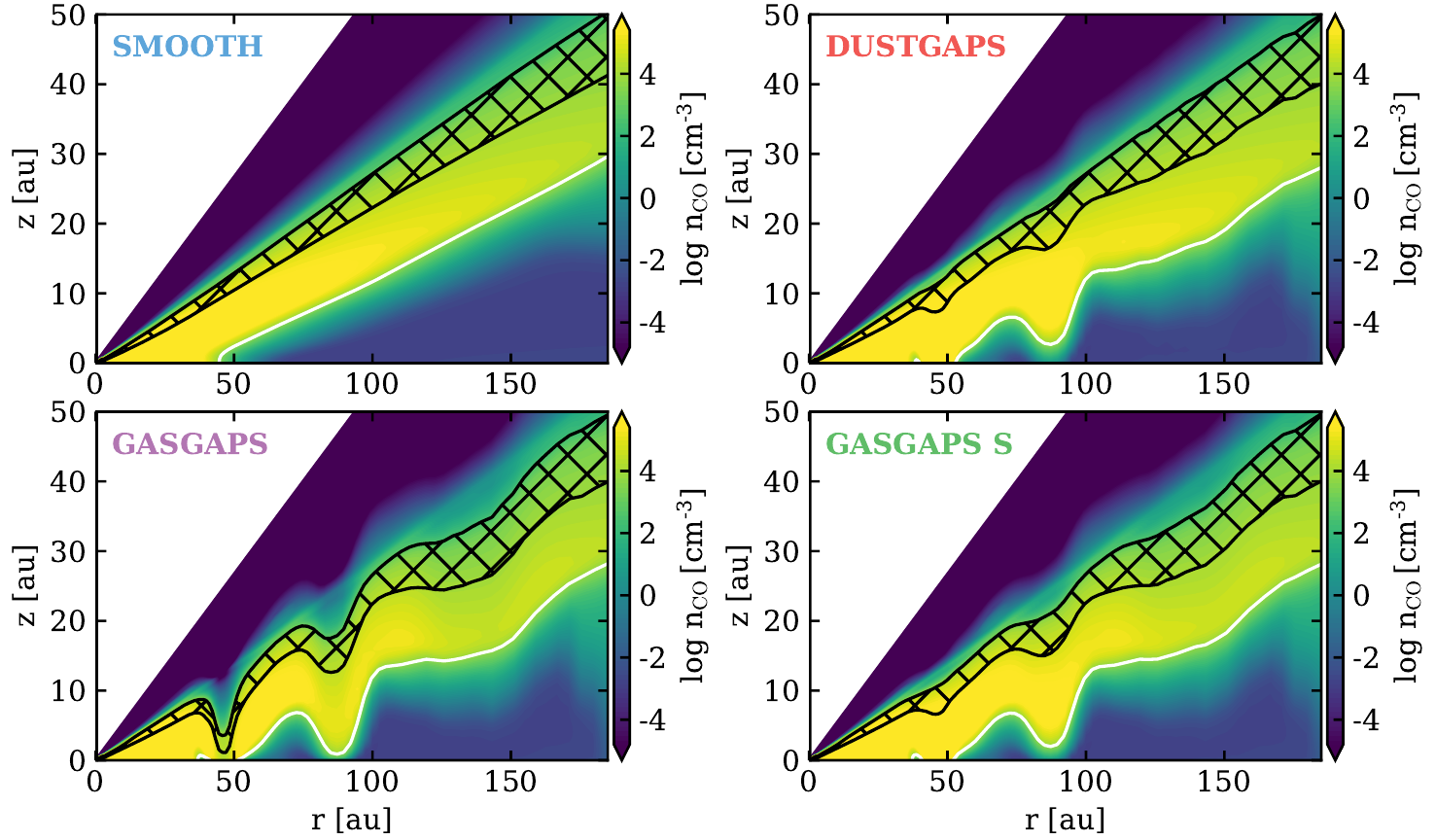}
\caption{CO number density and the \mbox{$^{12}\mathrm{CO}\,J\!=\!2\!-\!1$} emitting layer (hatched area) for the four  fiducial models. The upper and lower bound of the emitting layer at each radius correspond to 15 and 85 percent of the total flux, integrated from top to bottom. The white solid contour shows $T_\mathrm{dust}=25\,\mathrm{K}$. To determine the emission layer we assume that we are looking face-on onto the disk.}
\label{fig:lineorigin}
\end{figure*}
\section{Impact of thermo-chemical processes on the measured rotation velocity}
\label{sec:vrot}
\citet{Teague2018} use ALMA CO isotopologue observations to map the velocity field in the disk of \mbox{HD 163296}. They found deviations from the expected rotation velocity at the position of the observed dust gaps, and concluded that such deviations are likely caused by planets. Planets open up gaps in the gas and therefore affect the radial pressure gradients and consequently the measured rotation velocity.

If radial pressure gradients and the vertical extent of the disk are considered, the rotation velocity of a Keplerian disk is given by \citep[e.g.][]{Weidenschilling1977,Rosenfeld2013}
\begin{equation}
\frac{v_{\mathrm{rot}}^2}{r}=\frac{G M_*}{(r^2+z^2)^{3/2}}+\frac{1}{\rho}\frac{\partial P}{\partial r},
\label{eq:vrot}
\end{equation}
where $v_{\mathrm{rot}}$ is the rotation velocity, $G$ the gravitational constant, $M_*$ the mass of the star, $r,z$ are the radial and vertical spatial coordinates, $\rho$ the gas density, and $P$ the local gas pressure. We neglect here the impact of self-gravity since \citet{Rosenfeld2013} found that this term is of less importance. The second term of Eq.~(\ref{eq:vrot}) was used by \citet{Teague2018} to infer radial density gradients and profiles. However, they neglected the impact of temperature on the pressure gradient as their observations, in particular changes in the line width across the gaps,  suggest that the temperature is not the dominating factor. 

With the model presented here, for the first time we can study the impact of the temperature on $v_\mathrm{rot}$ in a self-consistent manner. To do this we use a similar presentation as \citet{Teague2018} but measure $v_\mathrm{rot}$ directly in the model instead of the observables. In this way we avoid problems arising from observational biases and limitations (e.g. spatial resolution).

\begin{figure}
\includegraphics{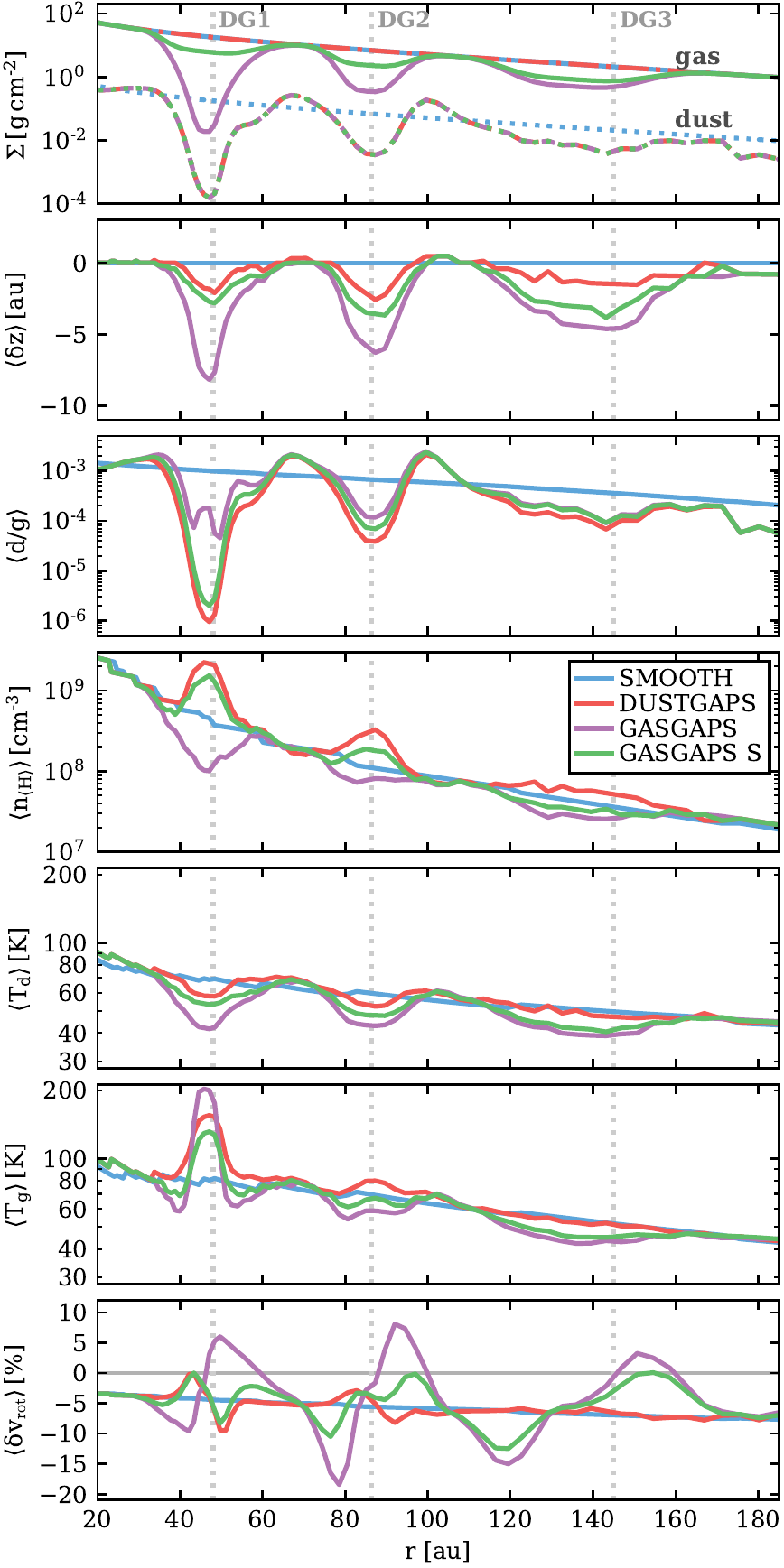}
\caption{Physical properties in the line emitting region of the \mbox{$^{12}\mathrm{CO}\,J\!=\!2\!-\!1$} line for the four fiducial models (see legend). The top row shows the gas and dust surface densities $\Sigma$ (all models with gaps have the same dust surface density profile). The other panels show quantities averaged over the line emitting region (from top to bottom):  shift in the height of the line emitting region $\langle \delta z \rangle$;  dust to gas mass ratio $\langle d/g \rangle$; total hydrogen number density $\langle n_\mathrm{\langle H\rangle} \rangle$; dust temperature $\langle T_\mathrm{d} \rangle$; gas temperature $\langle T_\mathrm{g} \rangle;$ and the change in the rotational velocity $\langle \delta v_\mathrm{rot} \rangle$ (see Eq.~\ref{eq:dvrot}). The vertical dotted lines indicate the locations of the dust gaps DG1, DG2, and DG2.}
\label{fig:vrot12CO}
\end{figure}
The line emitting region of the \mbox{$^{12}\mathrm{CO}\,J\!=\!2\!-\!1$} line in the models is shown in Fig.~\ref{fig:lineorigin}. At each radial grid point we integrate the line emission vertically from the top towards the disk midplane and define the top and bottom border of the region where the flux reaches 15\% and 85\% of the total flux, respectively. Within this line emitting layer we calculate several vertically averaged quantities as a function of radius, which are shown in Fig.~\ref{fig:vrot12CO}. The quantity $\langle \delta z \rangle$ is the change in the height of the line emitting region relative to the SMOOTH model;  $\langle d/g \rangle$ is the dust to gas mass ratio; $\langle n_\mathrm{\langle H \rangle} \rangle$ is the total hydrogen number density ($n_\mathrm{\langle H \rangle}\!=\!n_\mathrm{H}\!+\! 2 n_\mathrm{H_2}$); and $\langle T_\mathrm{d} \rangle$ and $\langle T_\mathrm{g} \rangle $ are the average dust and gas temperatures, respectively. Furthermore, we calculate the average rotation velocity $\langle v_\mathrm{rot}\rangle$ by using Eq.~(\ref{eq:vrot}) and vertically average $v_\mathrm{rot}$ in the same manner as the other quantities. For the presentation in Fig.~\ref{eq:vrot} we use $\langle\delta v_\mathrm{rot}\rangle,$ which we define, following \citet{Teague2018}, as the deviation from the expected Keplerian velocity $v_\mathrm{Kep}$ (i.e. neglecting the disk height and pressure gradients) 
\begin{equation}
\langle \delta v_\mathrm{rot} \rangle=\frac{\langle v_\mathrm{rot}\rangle-v_\mathrm{Kep}}{v_\mathrm{Kep}}\times 100 \;\;\;[\%].
\label{eq:dvrot}
\end{equation}
We also applied the above-described procedure for the \mbox{$\mathrm{C^{18}O}\,J\!=2\!-\!1$} line, which is shown in Fig.~\ref{fig:vrotC18O}. However, as the results are qualitatively very similar to the \mbox{$^{12}\mathrm{CO}\,J\!=\!2\!-\!1$} line, here we only discuss the results for the \mbox{$^{12}\mathrm{CO}\,J\!=\!2\!-\!1$} line.   
\subsection{Line emitting layer}
The height of the line emitting layer changes by up to \mbox{$\langle \delta z\rangle \approx -2\,\mathrm{au}$} (see Figures \ref{fig:lineorigin} and \ref{fig:vrot12CO}), when dust gaps are present (i.e SMOOTH versus DUSTGAPS model). The reason is the reduced optical depth within the dust gaps, which increases the temperature but also causes additional photo-dissociation of CO in the upper layers. The impact of the latter effect is however limited as CO is efficiently self-shielding within the gaps (see also \citealt{Facchini2018}). The change in temperature also affects the population levels of the CO molecule. In regions where the temperature increases, higher levels are populated, the \mbox{$^{12}\mathrm{CO}\,J\!=\!2\!-\!1$} line intensity decreases in those layers, and the line emitting layer of \mbox{$^{12}\mathrm{CO}\,J\!=\!2\!-\!1$} moves deeper into the disks with cooler temperatures and higher densities.

In the GASGAPS model the line emitting layer moves even deeper into the disk as now the gas density also drops within the dust gaps. The emitting layer within DG1 almost reaches the disk midplane. For this gap, $\Sigma_{g}$ was reduced by a factor of $1000$ with respect to the SMOOTH model.   In the model with shallow gas gaps (\mbox{GASGAPS~S}) the line emitting layer also moves deeper into the disk at the location of the dust gaps with respect to the DUSTGAPS model.  
\subsection{Rotation velocity}
For $\langle\delta v_\mathrm{rot}\rangle$ (bottom panel of Fig.~\ref{fig:vrot12CO}) we first discuss some general properties and then focus on the behaviour of $\langle\delta v_\mathrm{rot}\rangle$ around DG2 and DG3. Finally we discuss DG1, which shows significantly different physical properties from the other two gaps. 

As seen from Eq.~\ref{eq:vrot}, $v_\mathrm{rot}$ is sensitive to changes in the height of the emission layer and changes in the pressure gradient, which is affected by temperature and density. For the SMOOTH model $\langle\delta v_\mathrm{rot}\rangle<0$ at all radii because the emission layer is high above the midplane (Fig.~\ref{fig:lineorigin}) and the global radial surface density gradient used in our model is negative. The contribution of the latter to $\langle\delta v_\mathrm{rot}\rangle$ is about three percentage points at $r=180\,\mathrm{au}$, and becomes negligible for $r<100\,\mathrm{au}$ (see also \citealt{Pinte2018a}). Self-gravity would enhance $v_\mathrm{rot}$ and therefore would slightly shift $\langle\delta v_\mathrm{rot}\rangle$ towards zero. However, as already noted we neglect the self-gravity term here because it is of less importance \citep[see][]{Rosenfeld2013} and would not significantly affect our conclusions. 

Around the location of DG2 and DG3 we see a distinct pattern in $v_\mathrm{rot}$ in the models with gas gaps. This pattern is caused by the radial change of the pressure gradient, and is qualitatively similar to what was derived by \citet{Teague2018} from the pre-DSHARP data of \mbox{\mbox{HD 163296}}. Across a gap (in radial direction) the density first decreases, reaches its minimum, and then increases again, which results in a negative pressure gradient ($\langle\delta v_\mathrm{rot}\rangle< 0$), followed by a positive one ($\langle\delta v_\mathrm{rot}\rangle>0$) \citep{Teague2018}.

In the DUSTGAPS model we see an inverse pattern at DG2 compared to the models with a gas gap. As there is no gas depletion within the dust gap, this pattern is solely caused by the temperature gradient within DG2. As the gas temperature increases in the gap, due to dust depletion, the sign of the $\langle\delta v_\mathrm{rot}\rangle$ pattern is switched compared to the case of a gas gap. With respect to the SMOOTH model, $\langle\delta v_\mathrm{rot}\rangle$ is at most $\pm3\%$ in the DUSTGAPS model around DG2. 

For DG1 the gas temperature changes by about a factor of two within the dust gap, which is significantly higher than for DG2 and DG3. Due to the strong dust depletion that results in $\langle d/g \rangle<10^{-4}$ (depending on the gas depletion), the cooling of the gas via thermal accommodation on grains becomes less efficient. This is even stronger if the gas is also depleted as the cooling rate is proportional to the dust and gas number density \citep{Woitke2009a}. Furthermore, photo-electric heating  via  PAHs\footnote{PAHs are coupled to the gas in our model.} is more efficient inside the gap due to the enhanced far-UV field (see e.g. Fig.~\ref{fig:struclinear_DUSTGAPS}). For DG1 in the GASGAPS model viscous heating also kicks in, as the line emitting layer is close to the midplane, resulting in even higher gas temperatures.  

As a consequence of the gas temperature increase in DG1, the $\langle\delta v_\mathrm{rot}\rangle$ pattern is dominated by the temperature gradient in the \mbox{GASGAPS S} model. Only in the GASGAPS model, where the gas is depleted by a factor of 1000 in DG1, do we see the expected $\langle\delta v_\mathrm{rot}\rangle$ pattern for a gas gap but with weaker amplitudes (i.e. the gas gap appears less deep). As already noted by \citet{Teague2018} and \citet{vanderMarel2018}, a temperature increase within the gap will influence the derived gas surface density profile and also the $\langle\delta v_\mathrm{rot}\rangle$ pattern. This will lead to an underestimation of the mass of the forming planet that has opened up the gap. However, our models indicate that in the case of a deep dust gap, such as DG1 in \mbox{HD 163296}, the $\langle\delta v_\mathrm{rot}\rangle$ pattern can even be completely dominated by the temperature gradient and hide the presence of a gas gap.

\begin{figure}
\resizebox{\hsize}{!}{\includegraphics{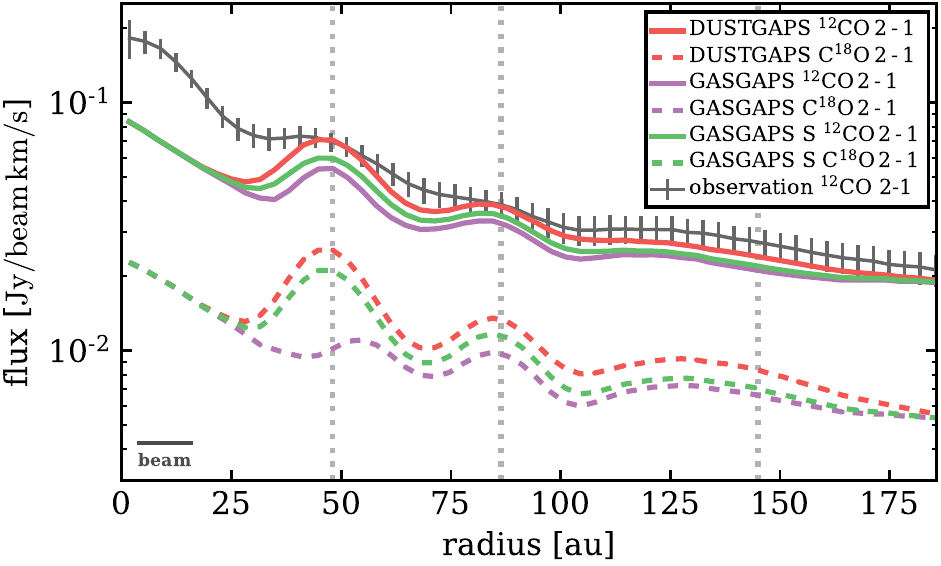}}
\caption{Azimuthally averaged radial intensity profiles for the \mbox{$^{12}\mathrm{CO}\,J\!=\!2\!-\!1$} (coloured solid lines) and  \mbox{$\mathrm{C^{18}O}\,J\!=\!2\!-\!1$} (dashed lines) spectral lines. The models shown are indicated in the legend. The grey solid line with error bars are the \mbox{$^{12}\mathrm{CO}\,J\!=\!2\!-\!1$} line observations. }
\label{fig:gasgapsC18O}
\end{figure}

We note that it is possible that we over-predict the temperature increase in DG1 in our models. This can be seen in the radial line intensity profiles where the peak at DG1 is too pronounced (see Sect.~\ref{sec:radprofs} and also Fig.~\ref{fig:gasgapsC18O}). The measurements of the \mbox{$^{12}\mathrm{CO}\,J\!=\!2\!-\!1$} brightness temperature by \citet{Isella2018} and \citet{Teague2019} seem to show a peak at the location of DG1 but not as strong as in our model. A reason for a too high temperature in the models could be our dust model (see Sect.~\ref{sec:radprofs}). We make the dust gaps by removing dust grains of all sizes, including the small ones (i.e. $< 1\,\mathrm{\mu m}$). This leads to less efficient gas cooling via thermal accommodation on grains and to a strong reduction of the optical depth at far-UV wavelengths and consequently to more efficient gas heating. As already noted in Sect.~\ref{sec:radprofs}, an improved dust model is required (i.e. radial variation of grain size distribution) to improve our fit to the data and to derive a more accurate temperature structure (see also Appendix ~\ref{sec:compFacc} for a comparison to the models of \citealt{Facchini2018}). Nevertheless, our models indicate that in deep dust gaps such as DG1 the impact of the temperature on $\langle\delta v_\mathrm{rot}\rangle$ is significant, contrary to the two shallow dust gaps.  

\begin{figure}
\includegraphics{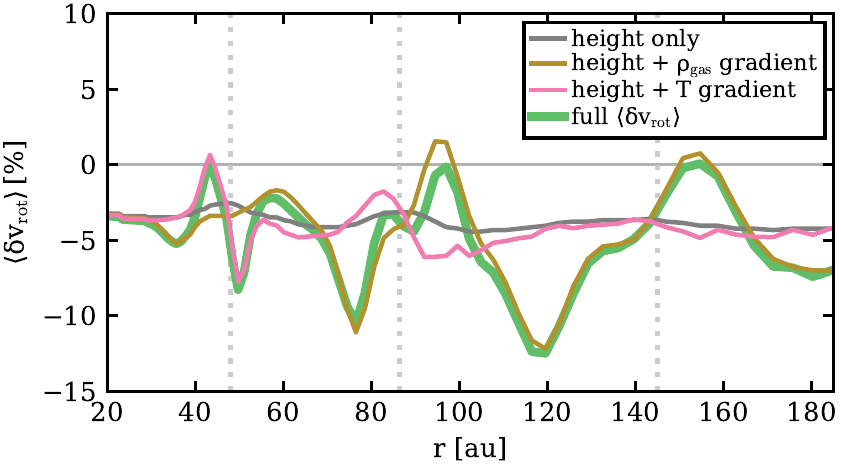}
\caption{Deviations $\langle \delta v_\mathrm{rot} \rangle$ from the Keplerian velocity for the \mbox{GASGAPS S} model. Here we show the individual contributions to $\langle \delta v_\mathrm{rot} \rangle$ from the height of the \mbox{$^{12}\mathrm{CO}\,J\!=\!2\!-\!1$} emission layer (grey solid line), the density gradient (brown solid line), and the temperature gradient (pink solid line). The green solid line shows $\langle \delta v_\mathrm{rot} \rangle$ including all the components (same as in the bottom panel of Fig.~\ref{fig:vrot12CO}).}
\label{fig:vrot12COGASGAPSS}
\end{figure}
In Fig.~\ref{fig:vrot12COGASGAPSS} we present again $\langle\delta v_\mathrm{rot}\rangle$ for the \mbox{GASGAPS S} model, but now showing the individual components that determine the $\langle\delta v_\mathrm{rot}\rangle$ profile. We note that for all three gas gaps in the \mbox{GASGAPS S} model the gas is depleted by a factor of three with respect to the SMOOTH model (see Sect.~\ref{sec:method_gaps}). Figure~\ref{fig:vrot12COGASGAPSS} clearly shows that only for DG3, the $\langle\delta v_\mathrm{rot}\rangle$ profile is fully determined by the density gradient. For DG2 the temperature gradient has some limited impact on $\langle\delta v_\mathrm{rot}\rangle$, but for DG1 the $\langle\delta v_\mathrm{rot}\rangle$ profile is mostly shaped by the temperature gradient. This figure demonstrates the complexity of $\langle\delta v_\mathrm{rot}\rangle$ measurements but also that each dust gap is quite unique with respect to the $\langle\delta v_\mathrm{rot}\rangle$ profile. 

Our analysis in this section shows the complex interplay of dust and gas gaps and thermo-chemical processes that all have an impact on the measured rotational velocity. Nevertheless, our analysis confirms the interpretation of \citet{Teague2018} that the measured deviations $\delta v_\mathrm{rot}$ for DG2 and DG3 in \mbox{HD 163296} are most likely caused by gas gaps. For DG1 the quality of the data used by \citet{Teague2018} did not allow for a firm conclusion. However, our dust model for \mbox{HD 163296} shows that the properties of DG1 are quite different to DG2 and DG3, in particular it is significantly deeper (see also \citealt{Isella2016}). DG1 is also the only dust gap in \mbox{\mbox{HD 163296}} that was detected in scattered light observations with VLT/SPHERE \citep{Muro-Arena2018}. The extreme dust depletion in DG1 makes the interpretation of gas line observations around DG1 more complex as our modelling shows. To fully understand the nature and origin of DG1, molecular line observations with a similar spatial resolution as the DSHARP continuum observations would be beneficial.%
\section{Potential gas gaps}
\label{sec:gasgaps}
Our analysis in Sect.~\ref{sec:vrot} has shown that both dust and gas gaps influence the \mbox{$^{12}\mathrm{CO}\,J\!=\!2\!-\!1$} line excitation and emitting region and the measured rotation velocity. In this section we discuss to what extent it is possible to infer the presence of gas gaps from the radial intensity profiles and channel maps of our models. Furthermore we also discuss our results with respect to previous gas gap modelling of \mbox{HD 163296}  from the literature. 

We note that we did not attempt to precisely fit the \mbox{$^{12}\mathrm{CO}\,J\!=\!2\!-\!1$} radial intensity profiles or the channel maps by modulating the gas density. First of all the \mbox{$^{12}\mathrm{CO}\,J\!=\!2\!-\!1$} is highly optically thick and therefore mostly insensitive to the gas density, and secondly 
as discussed in Sect.~\ref{sec:backside} the limitations of our dust model do not allow for accurate fitting of the gas lines. Nevertheless, our models and synthetic observables are still instructive to devise a strategy for interpreting real observations.%
\subsection{Radial intensity profiles}
In Fig.~\ref{fig:gasgapsC18O} we show the \mbox{$^{12}\mathrm{CO}\,J\!=\!2\!-\!1$} and \mbox{$\mathrm{C^{18}O}\,J\!=\!2\!-\!1$} radial intensity profiles for our models with gaps. For none of the models is a clear signature of a gas gap, namely a drop in intensity at the position of the dust gaps, visible. This is simply because of the high disk mass $M_\mathrm{disk}\sim0.2\,M_\odot,$ which makes even the \mbox{$\mathrm{C^{18}O}\,J\!=\!2\!-\!1$} line optically thick in \mbox{HD 163296} (see also \citealt{Booth2019}). Although in the GASGAPS model \mbox{$\mathrm{C^{18}O}\,J\!=\!2\!-\!1$} becomes optically thin in the centre of DG1 (measured along the z-axis of the disk), the inclination of the disk and the limited spatial resolution  prevents a direct view into the gap. If we view the model face-on we indeed see, at least in the GASGAPS model, a decrease in the line intensity across the dust gaps, although for \mbox{$^{12}\mathrm{CO}\,J\!=\!2\!-\!1$} we only see this in the two outermost gaps (see Fig.~\ref{fig:gasgapsC18OF}). This shows the limitation of azimuthally averaged radial intensity profiles for measuring gas gap properties even if the disk inclination is considered by de-projecting the data.
 
\begin{figure*}
\centering
\includegraphics{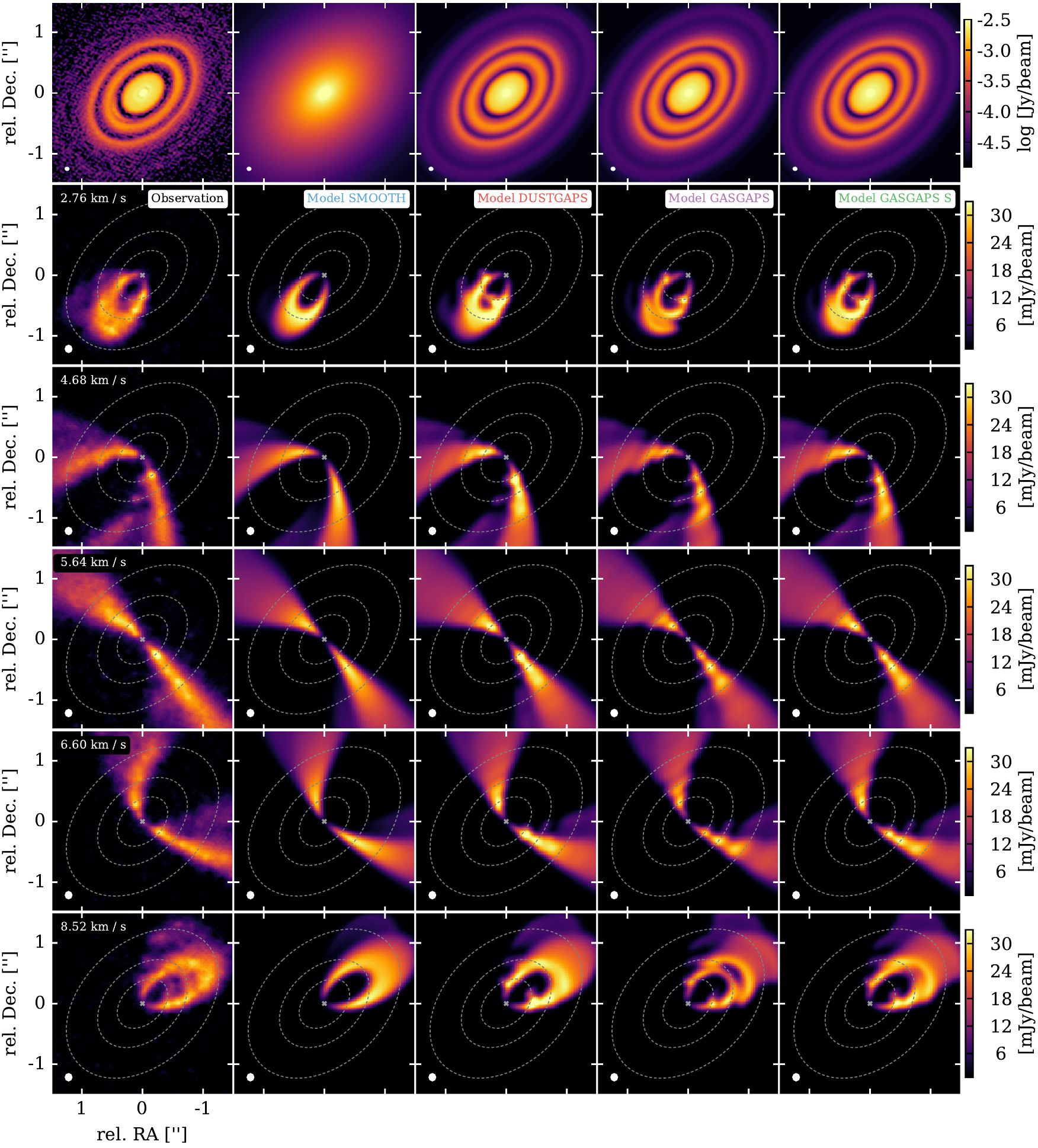}%
\caption{Dust continuum images and channel maps for the model series in comparison to the observations. Each row shows, from left to right, the observations, the SMOOTH, the DUSTGAPS, the GASGAPS, and the GASGAPS~S model. In the top row the dust continuum (note the logarithmic scale) is shown for comparison; the five remaining rows show the \mbox{$^{12}\mathrm{CO}\,J\!=\!2\!-\!1$} emission for selected velocity channels. The velocity is given in the top left corner of each row. The dashed ellipses  in the channel maps indicate the location of the dust gaps for reference. The filled white ellipses in the bottom left corner of  each panel indicate the beam size of the observations.}
\label{fig:channelmapsGG}
\end{figure*}
\subsection{Channel maps}
In Fig.~\ref{fig:channelmapsGG} we show five selected velocity channels of the DSHARP \mbox{$^{12}\mathrm{CO}\,J\!=\!2\!-\!1$} line observations in comparison to our model results. The modelled line cube was convolved with the same beam size as the observations, however we did not include the noise in the synthetic observables to show  the differences between the models more clearly, and also because of the high quality of the data. The goal here is not an exact comparison to the observations but rather to investigate what signatures of dust and/or gas gaps become visible in  channel maps.

The channel maps for the SMOOTH model show the perfectly smooth butterfly pattern as expected for an azimuthally symmetric disk with a smooth density and temperature structure. In the DUSTGAPS model we see clearly the impact of the dust rings on the CO emission from the back side of the disk, which is also apparent in the observations \citep{Isella2018}. However,  the front side of the disk (the bright emission) also shows clear differences compared to the SMOOTH model. At the location of the dust gaps the emission is brighter, which is caused by the higher temperature within the gaps (see also Sect.~\ref{sec:radprofs}). It is also apparent that the shape of the butterfly is no longer completely smooth, especially close to the position of the dust gaps. As there are no changes in the gas density compared to the SMOOTH model, those changes in the observed velocity can only be caused by the change of height of the emission layer and the temperature gradient within the dust gaps (see Sect.~\ref{sec:vrot}). 

In the models with gas gaps the structure in the channel maps becomes more pronounced as expected from changes in the pressure gradient. Compared to the data it seems that the distortions in the channel maps are too pronounced in the GASGAPS model. However, both the DUSTGAPS and GASGAPS~S model appear to be roughly consistent with the data. Such distortions are also visible in the observed channel maps, but the spectral resolution is likely to be too low to unambiguously confirm their presence. Compared to the \mbox{$^{12}\mathrm{CO}\,J\!=\!2\!-\!1$} radial profiles, the channel maps of the models show the differences in the physical model much more clearly (gas gaps or not etc.). Considering the channel maps and the $\delta v_\mathrm{rot}$ profiles (Sect.~\ref{sec:vrot}), our model with shallow gas gaps (GASGAPS S) is the one most consistent with the data.
\subsection{Comparison to other models}
Several models are published in the literature that attempted to model the pre-DSHARP dust and CO isotopologue data published in \citet{Isella2016}. To our knowledge the DSHARP dust and \mbox{$^{12}\mathrm{CO}\,J\!=\!2\!-\!1$} data have not yet been modelled in detail with a dust radiative transfer or thermo-chemical code. The spatial resolution of this  pre-DSHARP data is about six times lower for the continuum and two times lower for the \mbox{$^{12}\mathrm{CO}\,J\!=\!2\!-\!1$} line compared to the DSHARP data. 

The most similar approach to our modelling was done by \citet{vanderMarel2018} with the radiation thermo-chemical disk code DALI (Dust And LInes, e.g. \citealt{Bruderer2012b,Bruderer2013a}). Similarly to our results, they found that the radial line intensity profiles can actually show peaks at the position of the dust gaps even if gas gaps are present. However, they concluded that this is a temperature effect, whereas our analysis in Sect.~\ref{sec:backside} shows that at least for the gaps DG2 and DG3 the peaks are likely to be a result of the line emission from the back side (see Sect.~\ref{sec:backside}). Another difference is that in their models with deep gas gaps (depletion factors of 10 to 100) the line intensity at the position of the dust gaps also drops. They use rectangular gaps for both the gas and the dust (meaning the density depletion factor is constant across the gaps) while we use Gaussian-shaped gap profiles. Furthermore, there are differences in the underlying global disk structure and in the dust model (e.g. treatment of settling, dust opacities), which makes a detailed comparison difficult. However, both studies find the same physical effect of an increased radiation field and temperature in dust gaps. \mbox{\citet{vanderMarel2018}} concluded that with the data available at that time, it was not possible to confidently infer the presence of gas gaps in \mbox{HD 163296} and that the dust rings and gaps could also be caused by icelines, which do not require any gas gaps. In that case one would expect a $\langle \delta v_\mathrm{rot}\rangle$ pattern similar to our DUSTGAPS model (see Fig.~\ref{fig:vrot12CO}), but this pattern is not consistent with the observations of \cite{Teague2018}, at least not for DG2 and DG3.

\citet{Isella2016} also modelled the CO line data to derive gas depletion factors within the dust gaps. They used a model with a parameterized two-dimensional temperature structure and a simplified chemistry (e.g. fixed CO abundance and freeze-out at a certain temperature) and varied the CO surface densities for each line individually to fit the observational data. In their model the temperature does not change due to the presence of dust gaps or gas gaps, and also the dust and gas temperature are equal. For their narrow gap model (which is more consistent with the DSHARP data), they found gas depletion factors of 0-2.5, 30-70, and 3-6 for the gaps DG1 to DG3. For DG1 and DG3 this is consistent with our GASGAPS~S model, but the deep gas gap at the position of DG2 seems to be inconsistent with our results, in particular from the channel maps. \citet{Liu2018} used the same approach as \citet{Isella2016} for the temperature, chemical, and radiative transfer modelling but derived the gas surface density via hydrodynamic simulations including planets. In their best model the gas gap at the position of DG2 is rather shallow with a gas depletion factor of only a few and for DG1 they found a gas depletion factor of about ten.  However, their model is also still in reasonable agreement with the pre-DSHARP gas observations. In summary those models without thermo-chemistry seem to indicate the presence of gas gaps but the derived gas depletion factors can be quite different. The model of \citet{Liu2018} is roughly consistent with our model with shallow gas gaps (DUSTGAPS S, gas depletion factor of three in all gaps), and this is also our model that matches the DSHARP data best. 
\subsection{How to improve}
Deriving gas surface density profiles from observations is of particular importance to constrain hydrodynamic models and the origin of the gaps \citep[e.g.][]{vanderMarel2018}. If the gaps are produced by forming planets, it is also possible to derive planet masses from a comparison to hydrodynamic models. As an example we summarize here planet mass estimates that were derived for the DG2 in \mbox{HD 163296}. We have chosen DG2 because for that gap at least five different methods  were used in the literature to estimate the planet mass, including direct imaging. \citet{Zhang2018} estimated a planet mass in the range of $M_\mathrm{p}\sim0.07-0.6\,M_\mathrm{J}$ using the constraints from the DSHARP dust observations; using the pre-DSHARP dust and gas data \citet{Isella2016} and \citet{Liu2018} found $M_\mathrm{p}\sim0.3\,M_\mathrm{J}$ and $M_\mathrm{p}\sim0.46\,M_\mathrm{J}$, respectively; \citet{Teague2018} derived $M_\mathrm{p}\sim1\,M_\mathrm{J}$ by measuring pressure gradients and \citet{Pinte2020} reported (tentatively) $M_\mathrm{p}\sim1-3\,M_\mathrm{J}$ using the measured local deviations from Keplerian velocities in the \mbox{$^{12}\mathrm{CO}\,J\!=\!2\!-\!1$} DSHARP data. Taking the extreme values of the reported mass results in a range of $M_\mathrm{p}=0.07-3\,M_\mathrm{J}$, nearly two orders of magnitude. These mass estimates are consistent  with planet-mass upper limits  derived from high-contrast imaging with VLT/SPHERE ($M_\mathrm{p}=3-7 M_\mathrm{J}$; \citealt{Mesa2019}) and Keck/NIRC2 ($M_\mathrm{p}=4.5-6.5 M_\mathrm{J}$; \citealt{Guidi2018}).

As noted by \citet{Pinte2020}, planet mass measurements using constraints from gas observations tend to give higher planet masses than methods that only use constraints from dust observations. The methods using only gas observations neglect the impact of thermo-chemical processes. However, this most likely leads to an underestimation of the planet mass (see Sect.~\ref{sec:vrot}). Hydrodynamic models that use constraints from the dust observations suffer from a limited knowledge of the dust properties (e.g. sizes), the turbulence, or the disk scale-height. For example, the estimated planet mass can increase by about a factor of two for a turbulence alpha of $\alpha=10^{-3}$ compared to a case with $\alpha=10^{-4}$ \citep[e.g.][]{Zhang2018}. The upper limits derived from direct imaging depend on the system age and the planet and/or sub-stellar evolution model used \mbox{\citep[e.g][]{Mesa2019}}.

For \mbox{HD 163296} the available datasets are of exceptional quality but nevertheless inferring the presence of planets and in particular estimating their mass is still uncertain. As shown in this work, the interpretation of such data is very complex and the interplay of the dust and gas and thermo-chemical effects has to be considered. However, to improve this situation multi-wavelength data for the gas and the dust are also required. Multiple CO line transitions or observations of other molecules with a similar or better quality as the DSHARP data would provide much stronger constraints on the disk geometry (e.g. scale-height), temperature structure, and chemistry. For the dust, multi-wavelength observations and polarisation measurements can provide constraints on the radial variations of dust properties and also the disk geometry (e.g.~\citealt{Muro-Arena2018,Ohashi2019}). Only combined modelling of the dust and gas will allow us to unlock the full potential of such high-quality observations. 

For the models as presented here, more sophisticated dust modelling considering radial variations in the dust properties is required (for first attempts see e.g. \citealt{Akimkin2013,Facchini2017,Greenwood2019}). Such models will allow us to consider both dust and gas observations to constrain the radial dust properties and, by using various molecular tracers ,to eventually derive robust gas surface density profiles. Such information is crucial for hydrodynamic simulations and would provide the required input to improve the implementation of thermo-chemical processes in full hydrodynamic disk simulations \citep[for first attempts see e.g.][]{Vorobyov2020,Gressel2020}.%
\section{Conclusions}
\label{sec:conclusions}
We have presented a self-consistent thermo-chemical model to interpret the high spatial resolution DSHARP dust and gas data for the disk around the \mbox{Herbig Ae/Be} star \mbox{HD 163296}. We used the radiation thermo-chemical disk code P{\tiny RO}D{\tiny I}M{\tiny O} (PROtoplanetary DIsk MOdel) to investigate the importance of the dust component and thermo-chemical processes for the interpretation of high spatial resolution gas observations. We studied the impact of dust and gas gaps on observables such as radial intensity profiles and channel maps, and the role of thermo-chemical processes for measuring pressure gradients in disks. Our main findings are:
\begin{enumerate}
\item Azimuthally averaged radial \mbox{$^{12}\mathrm{CO}\,J\!=\!2\!-\!1$} intensity profiles derived from the DSHARP data clearly show peaks at the location of the three prominent dust gaps (DG1, DG2, and DG3) of \mbox{HD 163296} (Sect.~\ref{sec:radprofs}). The main contribution of those intensity peaks does not come from the increase of temperature within the dust gaps but from the emission of the back side of the disk. The line emission from the back side of the disk is partly absorbed by the dust rings but can pass through the dust gaps causing a bumpy radial \mbox{$^{12}\mathrm{CO}\,J\!=\!2\!-\!1$} intensity profile. A possible gas temperature increase within the dust gap can further enhance those peaks. This effect is not unique to \mbox{HD 163296} but will affect any millimetre-line observations including optically thin lines as long as the dust rings are at least marginally optically thick. A proper dust model is therefore crucial for interpreting high quality millimetre-line observations. 
\item We analysed the impact of thermo-chemical effects on pressure gradients and the measured rotational velocities $v_\mathrm{rot}$ (Sect.~\ref{sec:vrot}). We find that for deep dust gaps, such as DG1 at $r=48\,\mathrm{au}$ in \mbox{HD 163296}, temperature gradients within the gap can have a significant impact on the measured radial $v_\mathrm{rot}$  profile. For some cases the temperature gradient can even dominate the profile and hide the presence of any gas density gradient and gas gap. For the two other prominent dust gaps of \mbox{HD 163296} we find that the impact of temperature gradients is negligible and the measured $v_\mathrm{rot}$ profile  provides a good estimate of the density gradient and the presence of gas gaps. This is consistent with the findings of \citet{Teague2018} and confirms their  interpretation that those gaps are most likely carved by planets. However, the nature and origin of the deep dust gap DG1 (dust depletion factor of 100 to 1700) remains unclear, as the interpretation of the gas data for this gap is more complex. Higher spatial resolution gas data is required to draw a firm conclusion on the origin of that gap (e.g. one or multiple planets, dead zone, ice line).
\item We showed that azimuthally averaged radial line intensity profiles are difficult to interpret and are not sufficient to derive accurate gas surface density profiles for \mbox{HD 163296}. Both of our gas gap models with shallow gas gaps (gas depletion factor of a few) and deep gas gaps (depletion factor \mbox{$> 10$}) are roughly consistent with the observed \mbox{$^{12}\mathrm{CO}\,J\!=\!2\!-\!1$} radial intensity profile. However, a comparison of the models with the observed channel maps indicates that deep gas gaps are inconsistent with the data (Sect.~\ref{sec:gasgaps}). Also the model with only dust gaps produces distortions in the channel maps that seem to be consistent with observations. However, the dust gap only model would produce a distinctly different profile for the radial $v_\mathrm{rot}$ profile. This shows that considering the velocity information of gas line observations is crucial for the interpretation of dust and gas gaps in disks. 
\end{enumerate}
The existing high spatial resolution molecular line observations, which are mostly optically thick lines, and limitations in the models do not allow for the derivation of accurate gas surface density profiles. The large ALMA programme "The Chemistry of Planet Formation", will provide high spatial resolution data for various molecules, and the modelling will be further improved (e.g. in our case radially varying dust properties). Analysis of such data with models as presented here will provide more stringent constraints for hydrodynamic simulations and gap formation scenarios. It will also improve the mass estimates for forming planets and might tell us if there is a planet in each observed dust gap. Such information will be crucial to relate the forming planet population to the observed exoplanet population (see e.g. \citealt{Lodato2019,Ndugu2019}).  
\begin{acknowledgements}
We want to thank the anonymous referee for a very constructive report that improved the paper. C. R., G. M.-A., and C. G. acknowledge funding from the Netherlands Organisation for Scientific Research (NWO) TOP-1 grant as part of the research programme “Herbig Ae/Be stars, Rosetta stones for understanding the formation of planetary systems”, project number 614.001.552. This research has made use of NASA's Astrophysics Data System. This research made use of Astropy, a community-developed core Python package for Astronomy \citep{AstropyCollaboration2013}, matplotlib \citep{Hunter2007} and scipy \citep{2020SciPy-NMeth}. We would like to thank the Center for Information Technology of the University of Groningen for their support and for providing access to the Peregrine high performance computing cluster.

This paper makes use of the following ALMA data: ADS/JAO.ALMA\#2013.1.00366.S, ADS/JAO.ALMA\#2013.1.00601.S, and ADS/JAO.ALMA\#2016.1.00484.L. ALMA is a partnership of ESO (representing its member states), NSF (USA) and NINS (Japan), together with NRC (Canada), MOST and ASIAA (Taiwan), and KASI (Republic of Korea), in cooperation with the Republic of Chile. The Joint ALMA Observatory is operated by ESO, AUI/NRAO and NAOJ.
\end{acknowledgements}
\bibliographystyle{aa}
\bibliography{Rab_HD163296Gaps}
%
\begin{appendix} 
\section{Disk model}
\label{sec:diskmodel_app}
In Table ~\ref{table:diskmodel} we list the stellar and disk parameters for our model of \mbox{HD 163296}. These parameters are fixed for all models presented in the paper as we only vary the underlying gas and dust surface densities. However, we note that the gas to dust ratio changes as a function of radius and height depending on the presence of dust gaps and rings and because of dust settling. The used dust composition is a result of the \mbox{HD 163296} DIANA model \citep{Woitke2019}. To improve the match to the DSHARP observations, which were not included in the DIANA model, we optimized a few parameters by hand (e.g. $a_\mathrm{pow}$, $\alpha_\mathrm{set}$ , or $R_\mathrm{tap}$). However, the new model is still very similar to the DIANA model, and the agreement with the observational data used for the DIANA model is equally good.

The used settling prescription (Dubrulle settling; \citealt{Dubrulle1995}) depends on the local density (see \citealt{Woitke2016} for details) and therefore on the gas and dust depletion within the gaps. As a consequence the dust settling is slightly stronger in the models with gas gaps. However, this effect is not strong enough to affect our dust density profile significantly and the fit to the data is equally good for all models with dust gaps.   

For the vertical density structure we stick to the parameterized approach used for the DIANA model.  In this approach the scale height of the disk is given by  $H(r)=H(100\; \mathrm{au})\times(r/100\;\mathrm{au})^\beta$. We therefore do not self-consistently calculate the hydrostatic equilibrium structure.  However, this parameterized approach is a good approximation for the vertical hydrostatic equilibrium and is commonly used for thermo-chemical and radiative transfer modelling of disks \citep[e.g.][]{Muro-Arena2018,vanderMarel2018}. Furthermore, the parameterized approach makes the model calculations faster by factors of 10 to 100. Further details on the differences between parameterized and hydrostatic disk models are discussed in \citet{Woitke2016}.
\begin{table}
\caption{Main parameters for our reference HD~163296 model.}
\label{table:diskmodel}
\centering
\begin{tabular}{l|c|c}
\hline\hline
Quantity & Symbol & Value  \\
\hline
stellar mass \tablefootmark{a}                          & $M_\mathrm{*}$                    & $2~M_\sun$\\
stellar effective temp.               & $T_{\mathrm{*}}$                  & 9000~K\\
stellar luminosity                    & $L_{\mathrm{*}}$                  & $17~L_\sun$\\
\hline
strength of interst. FUV              & $\chi^\mathrm{ISM}$               & 1\tablefootmark{b}\\
\hline
min. dust particle radius             & $a_\mathrm{min}$                  & $\mathrm{0.02~\mu m}$\\
max. dust particle radius             & $a_\mathrm{max}$                  & 8.2 mm\\
dust size dist. power index           & $a_\mathrm{pow}$                  & 3.85 \\
turbulent mixing parameter      &  $\alpha_\mathrm{set}$                & $10^{-4}$ \\
max. hollow volume ratio\tablefootmark{c}              & $V_{\mathrm{hollow,max}}$         & 0.8\\
dust composition \tablefootmark{d}                     & {\small Mg$_{0.7}$Fe$_{0.3}$SiO$_3$}  & 69\%\\
(volume fractions)                    & {\small amorph. carbon}                    & 6\%\\
& {\small porosity}                          & 25\%\\
\hline 
\emph{main disk}  \\
disk gas mass                         & $M_{\mathrm{d}}$                  & $0.168~M_\mathrm{\sun}$\\
dust/gas mass ratio                   & $d/g$                             & 0.01\\
inner disk radius                     & $R_{\mathrm{in}}$                 & 2.5~au\\
tapering-off radius                   & $R_{\mathrm{tap}}$                & 110~au\\
column density power ind.             & $\epsilon$                        & 0.95\\
tapering-off exponent                 & $\gamma$                       & 0.7 \\
reference scale height                & $H(100\;\mathrm{au})$             & 7.55~au\\
flaring power index                   & $\beta$                           & 1.15\\
\hline
\emph{inner disk} \\
disk gas mass                         & $M_{\mathrm{d,I}}$                & $1.3\times10^{-4}~M_\mathrm{\sun}$\\
dust/gas mass ratio                   & $d/g_{I}$                             & $1.15\times10^{-5}$ \\
inner disk radius                     & $R_{\mathrm{in,I}}$                 & 0.41~au\\
outer disk radius                    & $R_{\mathrm{out,I}}$                & 2.5~au\\
column density power ind.             & $\epsilon_{I}$                        & 1.11\\
reference scale height                & $H_{I}(1\;\mathrm{au})$             & 0.077~au\\
flaring power index                   & $\beta_{I}$                           & 1.0\\
max. dust particle radius             & $a_\mathrm{max,I}$                  & $2.4\,\mathrm{\mu m}$\\
\hline
inclination\tablefootmark{e}          & $i$           & $46.7^\circ$       \\
position angle                        & $PA$          & $133.3^\circ$      \\
distance\tablefootmark{f}                              & $d$           & $101\,\mathrm{pc}$ \\
\hline
\end{tabular}
\tablefoot{
For more details on the parameter definitions, see \citet{Woitke2009a,Woitke2011,Woitke2016}.\\
\tablefoottext{a}{Stellar properties consistent with \citet{Fairlamb2015}.} \\
\tablefoottext{b}{$\chi^\mathrm{ISM}$ is given in units of the Draine field \citep{Draine1996b,Woitke2009a}.}\\
\tablefoottext{c}{Distributed hollow spheres dust opacities \citep{Min2005,Min2016}.} \\
\tablefoottext{d}{Optical constants from \citet{Dorschner1995a} and \citet{Zubko1996c}.}\\
\tablefoottext{e}{Inclination and position angle are from Table 2 of \citet{Huang2018}.}\\
\tablefoottext{f}{GAIA distance from \citet{Andrews2018}; consistent with \citet{Fairlamb2015}.}
}
\end{table}

\begin{figure*}
\centering
\includegraphics[width=\textwidth]{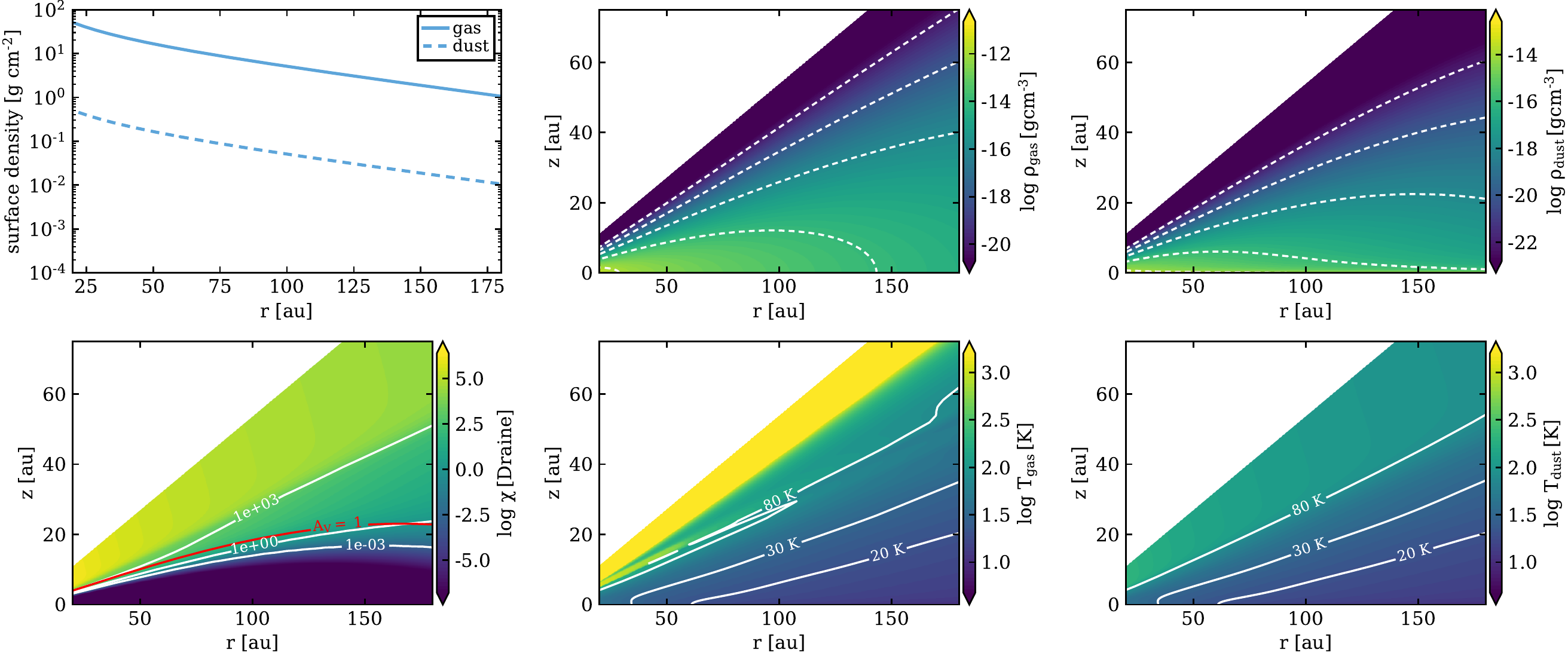}
\caption{Two-dimensional disk structure for the SMOOTH model. From top left to bottom right: Surface densities, gas density $\rho_\mathrm{gas}$, dust density ($\rho_\mathrm{dust}$), far-UV radiation field ($\chi$), gas temperature ($T_\mathrm{gas}$),  and dust temperature ($T_\mathrm{dust}$). The 2D contour plots only show the region from $r=20$ to $r=180\,\mathrm{au}$, to clearly show the physical properties within the prominent dust gaps (e.g. Fig.~\ref{fig:struclinear_DUSTGAPS}). The contour lines in the plots for the gas and dust density correspond to the values indicated in the  colour bar. The red contour line in the radiation field plot (first column bottom row) indicates a visual  extinction of unity.}
\label{fig:struclinear_SMOOTH}
\end{figure*}

\begin{figure*}
\centering
\includegraphics[width=\textwidth]{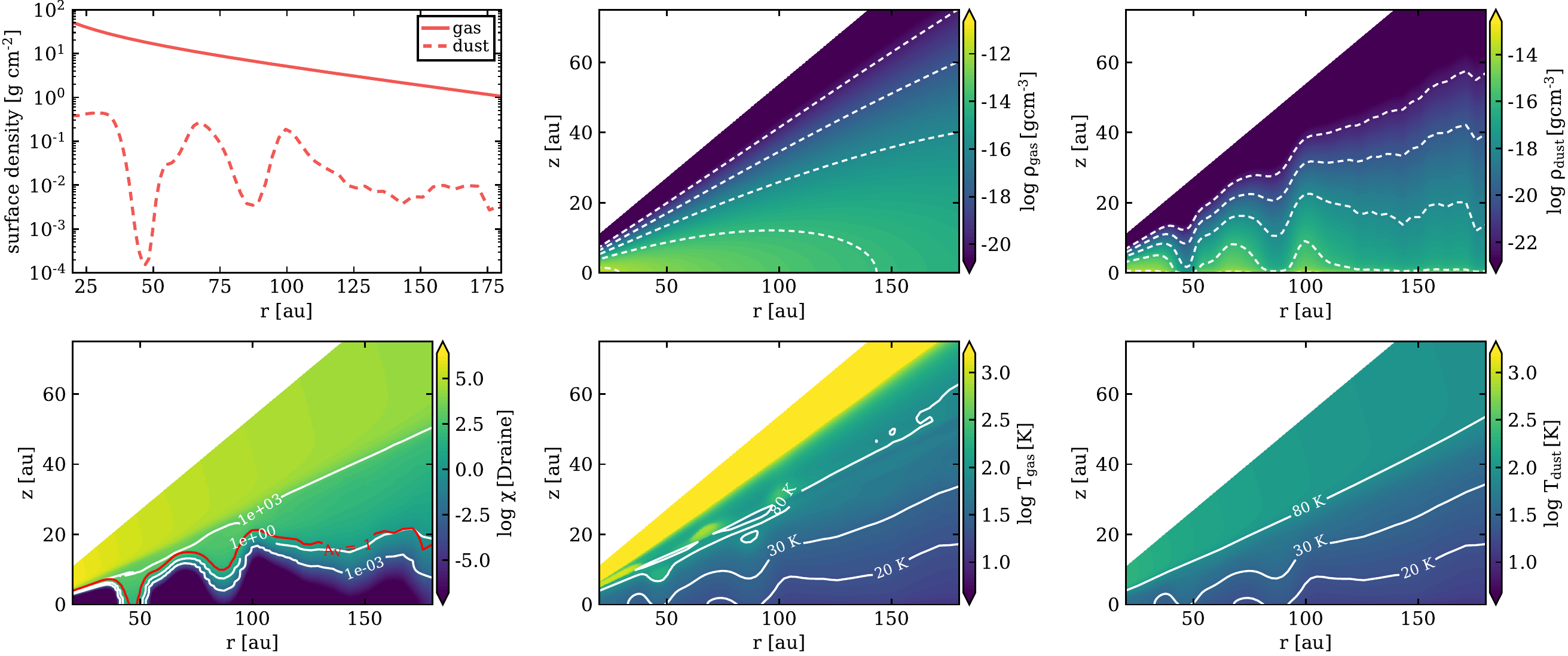}
\caption{Same as Fig.~\ref{fig:struclinear_SMOOTH} but for the DUSTGAPS model.}
\label{fig:struclinear_DUSTGAPS}
\end{figure*}

\begin{figure*}
\centering
\includegraphics[width=\textwidth]{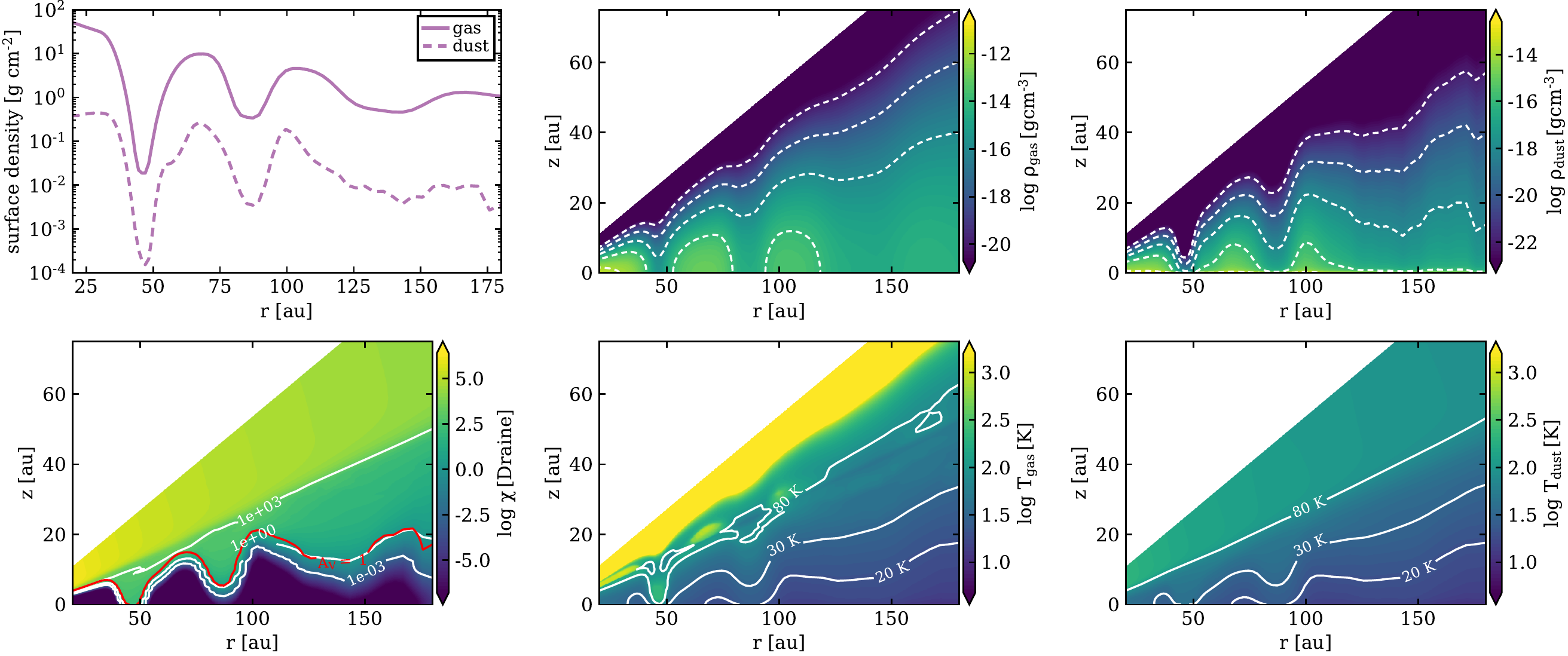}
\caption{Same as Fig.~\ref{fig:struclinear_SMOOTH} but for the GASGAPS model.}
\label{fig:struclinear_GASGAPS}
\end{figure*}

\begin{figure*}
\centering
\includegraphics[width=\textwidth]{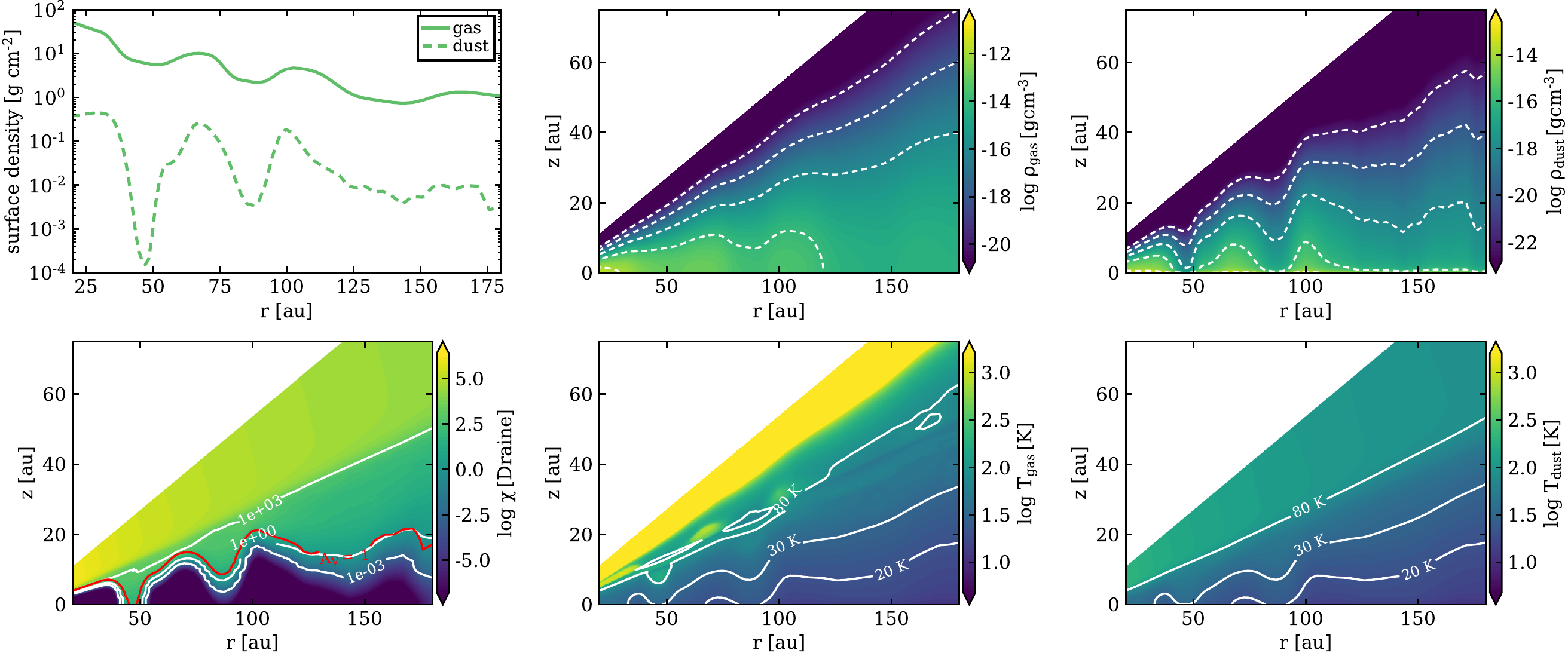}
\caption{Same as Fig.~\ref{fig:struclinear_SMOOTH} but for the GASGAPS~S model.}
\label{fig:struclinear_GASGAPS_S}
\end{figure*}

In Figures \ref{fig:struclinear_SMOOTH}, \ref{fig:struclinear_DUSTGAPS},  \ref{fig:struclinear_GASGAPS}, and \ref{fig:struclinear_GASGAPS_S}  we show the two-dimensional disk density structure, temperature  structure, and the radiation field for the SMOOTH, DUSTGAPS, GASGAPS, and \mbox{GASGAPS S} models. Also shown are the input surface densities for the gas and the dust for reference. A comparison of Figs.~\ref{fig:struclinear_SMOOTH} and \ref{fig:struclinear_DUSTGAPS} clearly shows the impact of the dust gaps on the radiation field and temperature structure. The visual extinction $A_\mathrm{V}$ decreases within dust gaps and the first dust gap becomes even optically thin ($A_\mathrm{V}<1$). As a consequence the dust and gas temperature are always higher within the dust gaps with respect to the regions around the gaps (the rings). 
\section{Separating the back and front side}
\label{sec:keplerianmasking}
\begin{figure}
\centering
\resizebox{\hsize}{!}{\includegraphics{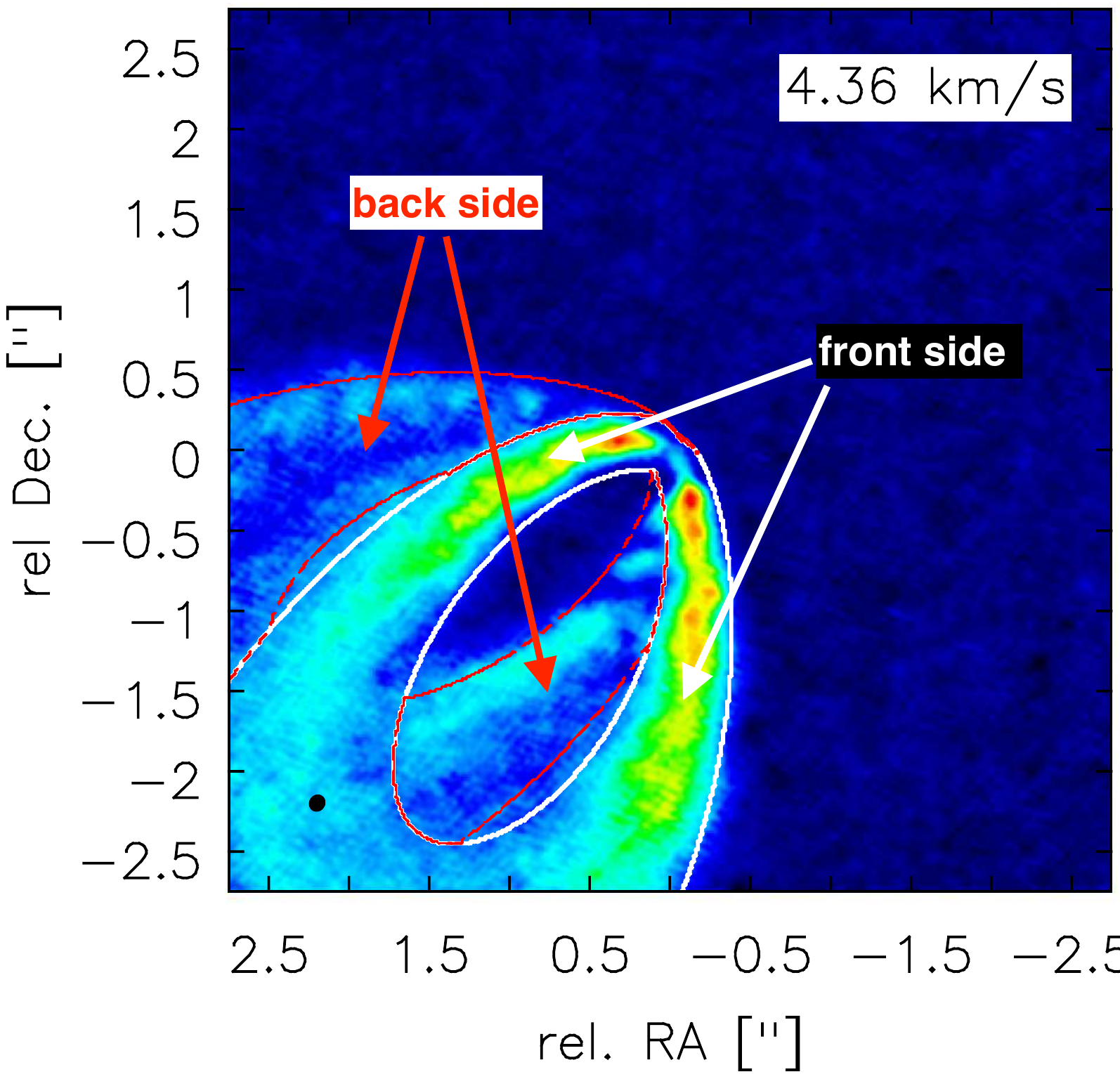}}
\caption{One velocity channel of the \mbox{$^{12}\mathrm{CO}\,J\!=\!2\!-\!1$} line cube showing the Keplerian masks for
the front (white solid line) and back side (red dashed line) of the disk. The colour scale  shows the line intensity in Jy/beam.}
\label{fig:masking}
\end{figure}

We used Keplerian masking to produce \mbox{$^{12}\mathrm{CO}\,J\!=\!2\!-\!1$} radial intensity profiles for the front and back side of the disk (see Fig.~\ref{fig:radmasks}). To construct the Keplerian masks we applied the \texttt{CLEAN\_mask} routine from the python package \texttt{imgcube}\footnote{\url{https://github.com/richteague/imgcube}} developed by \mbox{R. Teague}. As input for the routine we used the same values for the stellar mass, inclination, and position angle as for the models. The height of the emission layer was matched by eye by comparing the generated masks with the observational data. We slightly adapted the \texttt{imgcube} routine \texttt{CLEAN\_mask} to produce separate masks for the front and back side of the disk. To produce a mask for the visible part of the back side, we simply removed the areas of the back side mask where it overlaps with the front side mask. In Fig.~\ref{fig:masking} we show the two masks for one channel of the \mbox{$^{12}\mathrm{CO}\,J\!=\!2\!-\!1$} line cube. Using those masks we produced two line cubes one for the front side and one for the back side of the disk. For these line cubes we then applied the same procedure as for the full cube (see Sect.~\ref{sec:methods_obs}) to produce radial intensity profiles. 
\section{Comparison to \citet{Facchini2018}}
\label{sec:compFacc}
As noted in the main text, our model does not allow for radially varying dust properties (i.e. dust size distribution). This could have an impact on the physical properties within the gap. For example, if the small dust grains ($\lesssim 1\,\mathrm{\mu m}$) are not strongly depleted within a dust gap detected at millimetre wavelengths, they could efficiently shield the deeper regions of the gap from far ultraviolet (FUV) radiation. It is out of the scope of this paper to study this in detail but we want to compare our results concerning the physical and chemical conditions within the gaps to the thermo-chemical model of \citet{Facchini2018} that includes radially varying dust properties.

\citet{Facchini2018} modelled a disk around a solar-like star ($M_*=1\,M_\mathrm{\sun},  L_*=3\,L_\mathrm{\sun}$) with a gap at $r\sim20\,\mathrm{au}$. The gas density structure was given by hydrodynamic modelling including a planet (with different masses) that produced the gap in the gas density distribution. To determine the radial dust density structure, they used a 1D dust evolution code that considers a radially varying grain size distribution; for the vertical structure they included dust settling. This 2D dust and gas density structure is used as input for the radiation thermo-chemical disk code DALI (e.g. \citealt{Bruderer2012b,Bruderer2013a}) to determine the disk radiation field, gas and dust temperature,  and chemical abundances. We want to emphasize that the \citet{Facchini2018} model is not representative for \mbox{HD 163296} and their modelled gap is also much closer to the star ($r=20\,\mathrm{au}$) than the three prominent gaps for \mbox{HD 163296,} which are all located at $r\gtrsim 50\,\mathrm{au}$.

Similar to our results, they find that due to the presence of the dust gaps the FUV radiation can penetrate deeper into the disk resulting in a stronger FUV radiation field compared to the surrounding disk, and they also find that the dust temperature increases inside the gap. However, for the gas temperature, their model shows a decrease within the dust gap in the deeper layers of the disk. This is different to our models where the gas temperature always increases within dust gaps (see Figures \ref{fig:struclinear_DUSTGAPS}, \ref{fig:struclinear_GASGAPS}, and \ref{fig:struclinear_GASGAPS_S}). \citet{vanderMarel2018} also find that the gas temperature increases within the gaps in their model for \mbox{HD 163296}. They also used DALI, but with a simpler dust model than \citet{Facchini2018}. The reason for the difference in the gas temperature might be that for DG1 in our model, where we have the strongest temperature increase, the FUV radiation field within the gap is stronger by a factor of $\gtrsim 10$ than in the model of \citet{Facchini2018}. This could be caused by the different treatments of the small dust grain population, which can shield FUV radiation efficiently. However, we also note that in our model the stellar UV excess in the range of $6-13.6\,\mathrm{eV}$ is about three orders of magnitude higher than in the model of \citet{Facchini2018}, simply because of the different stellar properties used in the models.

According to \citet{Facchini2018} the decrease in gas temperature within the gap has a significant impact on the CO line emission, as most of the CO lines remain optically thick across the gap (similar to our model). Consequently, they find that their modelled gas gap is clearly visible in their synthetic CO line images, contrary to our models and the models of \citet{vanderMarel2018}. If the gas temperature decreased within the gaps of \mbox{HD 163296,} it would become even more difficult to reproduce the peaks in the observed radial \mbox{$^{12}\mathrm{CO}\,J\!=\!2\!-\!1$} intensity profile. The only way to produce peaks would be via the absorption of line emission from the back side of the disk by the dust (see Sect.~\ref{sec:backside}). However, it is unclear how strong this effect is in the model of \citet{Facchini2018} as they only present synthetic observables for the disk viewed face-on and also do not show channel maps. 

This comparison indicates again that the dust model is crucial for determining the gas properties. However, a more systematic approach of thermo-chemical dust and gas gap modelling in disks considering, for example, different disk masses, gaps at different radii, different dust models, and stellar properties is necessary to improve our understanding of high spatial resolution gas line observations and the physical properties of the gas within dust gaps.
\section{Radial profiles for the disk seen face-on}
\label{sec:faceon}
\begin{figure}
\resizebox{\hsize}{!}{\includegraphics{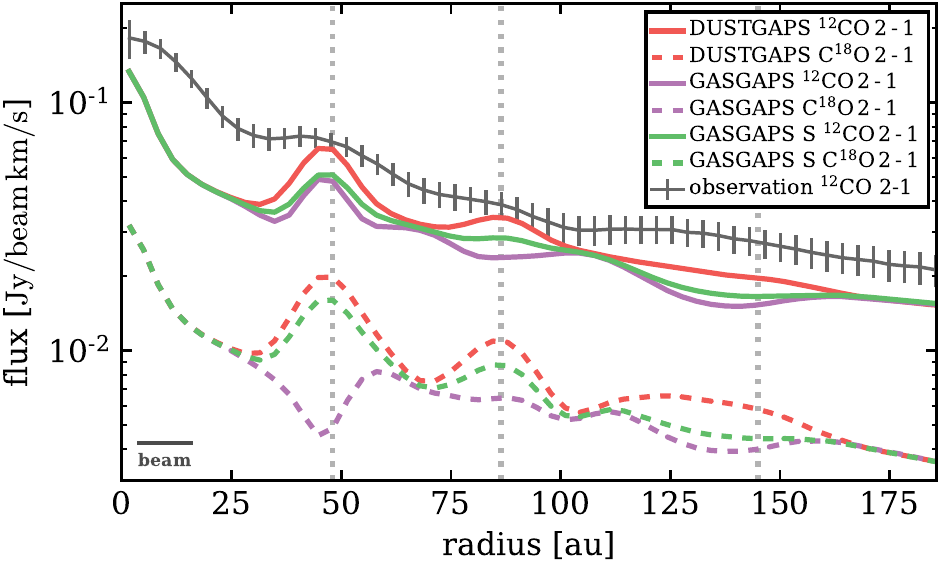}}
\caption{Same as Fig.~\ref{fig:gasgapsC18O} but the inclination for the models is $i=5^\circ$. The dotted vertical lines indicate the location of the three prominent dust gaps. The grey solid line with error bars shows the \mbox{$^{12}\mathrm{CO}\,J\!=\!2\!-\!1$} observations for reference.}
\label{fig:gasgapsC18OF}
\end{figure}

In Fig.~\ref{fig:gasgapsC18OF} we show the same models as in Fig.~\ref{fig:gasgapsC18O} but the disk is now seen nearly face-on ($i=5^\circ$). If the disk is seen face-on, emission from the back side of the disk will not be visible as both CO lines are optically thick. This is especially apparent for DG3 where the peak in the radial profile vanished in the \mbox{DUSTGAPS} model. For the other dust gaps the peak is still there but is less pronounced, as  we now see mostly the impact of the temperature increase within the dust gap. For DG3 the temperature change is not significant. 

For the models with gas gaps the emission of the \mbox{$^{12}\mathrm{CO}\,J\!=\!2\!-\!1$} line decreases in DG2 and DG3 for the face-on case, as we now have a more direct view into the dust gaps. Furthermore, at DG2 and DG3 the gas temperature is lower in the line-emitting region (see Fig.~\ref{fig:vrot12CO}) in the models with gas gaps compared to the \mbox{DUSTGAPS} model, hence the weaker emission. For DG1 the temperature increase is much stronger and we always see a peak in the \mbox{$^{12}\mathrm{CO}\,J\!=\!2\!-\!1$} radial profile at the location of DG1. For the \mbox{$\mathrm{C^{18}O}\,J\!=\!2\!-\!1$} line the situation is similar except for DG1 in the GASGAPS model. There the gas depletion within the gap is so strong that \mbox{$\mathrm{C^{18}O}\,J\!=\!2\!-\!1$} becomes optically thin, and the line emission drops significantly. 
\section{Radial properties for the \mbox{$\mathrm{C^{18}O}\,J\!=\!2\!-\!1$} line-emitting layer}
\label{sec:vrotC180}
\begin{figure}
\includegraphics{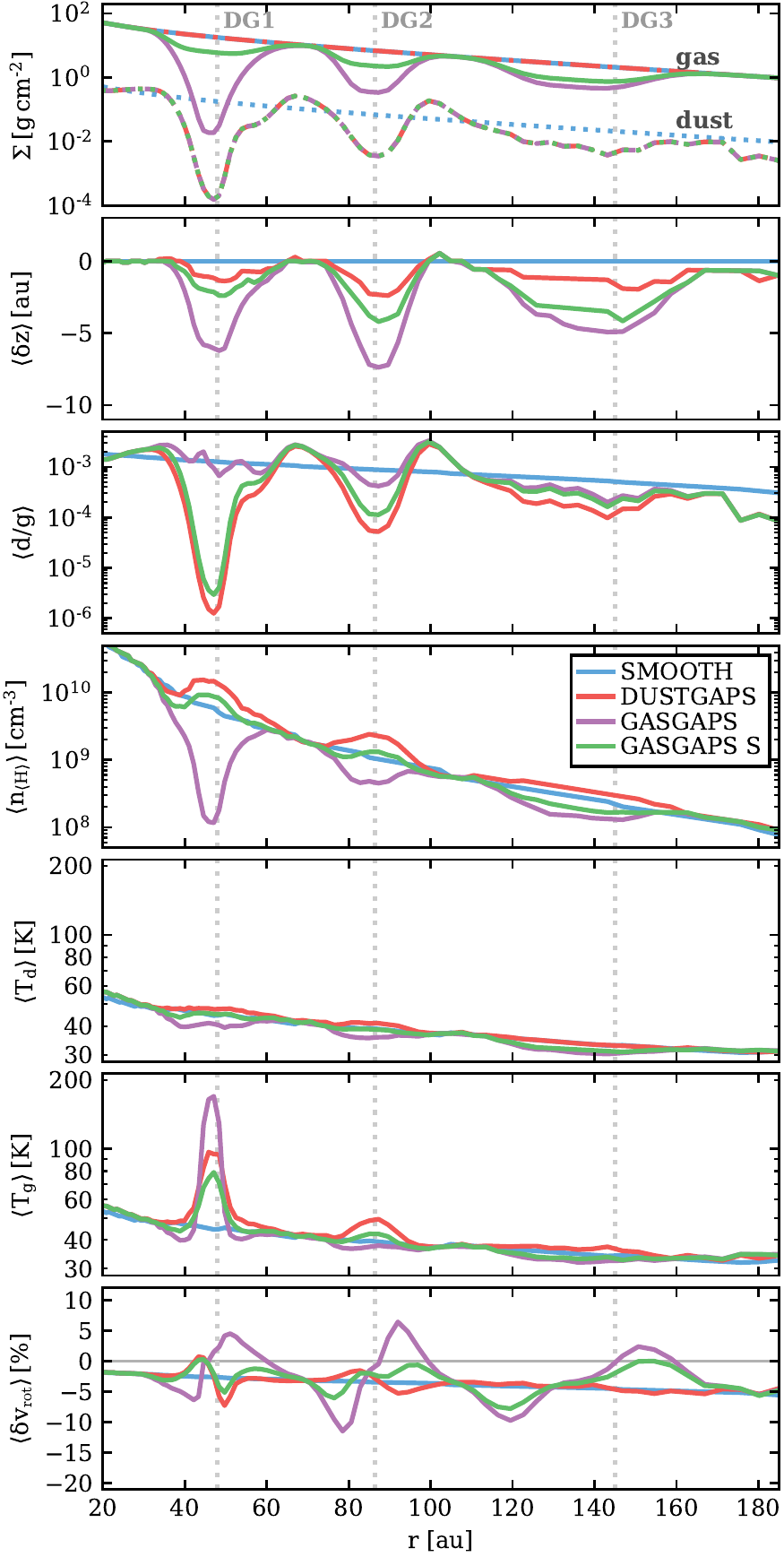}
\caption{Same as Fig.~\ref{fig:vrot12CO} but for the \mbox{$\mathrm{C^{18}O}\,J\!=\!2\!-\!1$} line.}
\label{fig:vrotC18O}
\end{figure}

In Fig.~\ref{fig:vrotC18O} we show the same averaged quantities as a function of radius  for the \mbox{$\mathrm{C^{18}O}\,J\!=\!2\!-\!1$} line emitting layer as are shown in Fig.~\ref{fig:vrot12CO} for the \mbox{$^{12}\mathrm{CO}\,J\!=\!2\!-\!1$} line (see Sect.~\ref{sec:vrot} for a detailed description). Although the absolute numbers are different compared to the \mbox{$^{12}\mathrm{CO}\,J\!=\!2\!-\!1$} line, the general picture is very similar. The $\langle \delta v_\mathrm{rot}\rangle$ profile shows the same pattern as for the \mbox{$^{12}\mathrm{CO}\,J\!=\!2\!-\!1$} line only with weaker amplitudes. This is expected as in our models the \mbox{$\mathrm{C^{18}O}\,J\!=\!2\!-\!1$} line is optically thick. Only for DG1 in the GASGAPS model, with a gas depletion factor of 1000, does the \mbox{$\mathrm{C^{18}O}\,J\!=\!2\!-\!1$} line become optically thin. 
\end{appendix}
\end{document}